\documentclass[11.5pt, twocolumn]{article}
\usepackage[a4paper, total={6.5in, 9.5in}]{geometry}
\pdfoutput=1

\usepackage{graphicx}
\usepackage{dcolumn}   
\usepackage{bm}        
\usepackage{amssymb}   
\usepackage{float}
\usepackage{xcolor}
\usepackage{soul}
\usepackage[normalem]{ulem}
\usepackage{authblk}
\usepackage{amsmath}
\usepackage{doi}
\usepackage{url}

\newif\ifcomments
\commentstrue

\ifcomments
  
  \newcommand{\strikeout}[1]{\st{#1}}

  \newcommand{\andresC}[1]{{\color{magenta}\textit{\textbf{Andr\'es:} #1}}}
  \newcommand{\matthiasC}[1]{{\color{red}\textit{\textbf{Matthias:} #1}}}
  \newcommand{\shimpeiC}[1]{{\color{red}\textit{\textbf{Shimpei:} #1}}}
\else

  \newcommand{\strikeout}[1]{}
  
  \newcommand{\andresC}[1]{}
  \newcommand{\matthiasC}[1]{}
  \newcommand{\shimpeiC}[1]{}
\fi

\newcommand{\psin}[0]{\psi_\mathrm{N}}
\newcommand{\grad}[0]{\bm\nabla}
\newcommand{\Podiv}[0]{P_\mathrm{outer\,\,div}}

\begin{document}

\title{Comparing spontaneous and pellet-triggered ELMs via non-linear extended MHD simulations}

\author[1,2]{\large A.~Cathey\thanks{andres.cathey@ipp.mpg.de}} 
\author[1]{\large M.~Hoelzl}
\author[3]{\large S.~Futatani}
\author[1]{\large P.T.~Lang}
\author[1]{\large K.~Lackner}
\author[4,5]{\large G.T.A.~Huijsmans} 
\author[6]{\large S.J.P.~Pamela} 
\author[1]{\large S.~G\"unter}
\author[7]{\large the JOREK team} 
\author[8]{\large the ASDEX Upgrade Team} 
\author[9]{\large the EUROfusion MST1 Team} 

\affil[1]{\small Max Planck Institute for Plasma Physics, Boltzmannstr.2, 85748 Garching, Germany}
\affil[2]{\small Physik Department, TUM, 85748 Garching, Germany}
\affil[3]{\small Universitat Politècnica de Catalunya, Barcelona, Spain}
\affil[4]{\small CEA, IRFM, 13108 Saint-Paul-Lez-Durance, France}
\affil[5]{\small Eindhoven University of Technology, P.O. Box 513, 5600 MB Eindhoven, The Netherlands}
\affil[6]{\small CCFE, Culham Science Centre, Abingdon, Oxon, OX14 3DB, United Kingdom}
\affil[7]{\small see https//www.jorek.eu for a list of current team members}
\affil[8]{\small see the author list of H. Meyer et al. 2019 Nucl. Fusion 59 112014}
\affil[9]{\small see the author list of B. Labit et al. 2019 Nucl. Fusion 59  0860020}

\date{}
\maketitle

\begin{abstract}
Injecting frozen deuterium pellets into an ELMy H-mode plasma is a well established scheme for triggering edge localized modes (ELMs) before they naturally occur. Based on an ASDEX Upgrade H-mode plasma, this article presents a comparison of extended MHD simulations of spontaneous type-I ELMs and pellet-triggered ELMs allowing to study their non-linear dynamics in detail. In particular, pellet-triggered ELMs are simulated by injecting deuterium pellets into different time points during the pedestal build-up described in [A.~Cathey et al.\ Nuclear Fusion 60, 124007 (2020)]. Realistic ExB and diamagnetic background plasma flows as well as the time dependent bootstrap current evolution are included during the build-up to capture the balance between stabilising and destabilising terms for the edge instabilities accurately. Dependencies on the pellet size and injection times are studied. The spatio-temporal structures of the modes and the resulting divertor heat fluxes are compared in detail between spontaneous and triggered ELMs. We observe that the premature excitation of ELMs by means of pellet injection is caused by a helical perturbation described by a toroidal mode number of ${\mathrm{n}=1}$. In accordance with experimental observations, the pellet-triggered ELMs show reduced thermal energy losses and narrower divertor wetted area with respect to spontaneous ELMs. The peak divertor energy fluency is seen to decrease when ELMs are triggered by pellets injected earlier during the pedestal build-up.
\end{abstract}

\section{Introduction}\label{intro}

Type-I edge localized modes (ELMs) are expected to induce large losses and consequently excessive transient divertor heat loads in large machines like ITER, which can affect the
lifetime of the components~\cite{Loarte2014}. Low-frequency large type-I ELMs must therefore be avoided in ITER either by small-ELM scenarios~\cite{Labit_2019}, ELM-free scenarios like QH-mode~\cite{Viezzer_2018}, active control via external resonant magnetic perturbations (RMPs)~\cite{Evans2008} or pellet ELM triggering~\cite{Lang2004_1,Baylor2013PRL}. Since RMPs may not be applicable in all phases of plasma operation (e.g., ramp-up and ramp-down) due to particular constraints on the plasma parameters and the edge safety factor and the possibilities of access to no-ELM or small-ELM regimes remain uncertain, pellet ELM triggering may offer a complementing approach and is foreseen as backup scheme in ITER. Additionally, the use of ELM pacing by means of pellet injection is an option for ITER to avoid the accumulation of impurities in the plasma~\cite{Loarte2014}. It has been shown experimentally that pellets allow to increase the ELM frequency and reduce the thermal energy losses associated to an individual ELM crash~\cite{Lang2004_1}. For pellet ELM triggering to be successful in mitigating the impact of ELMs on the divertor lifetime, the properties of spontaneous and pellet-triggered ELMs need to be investigated in direct comparison because experimental observations suggest that a reduced extent of the wetted area can cancel out the beneficial effects of the decreased energy expelled by pellet-triggered ELMs~\cite{Wenninger2011}. 

In the present article, directly comparable simulations of spontaneous and pellet-triggered ELMs are performed with the non-linear extended MHD code JOREK~\cite{Huysmans2007,Czarny2008,Hoelzl2021}. For this purpose, pellet injections are considered during different time points of the pedestal build-up; the setup is thoroughly described in Ref.~\cite{Cathey2020}. Realistic resistivity, heat diffusion anisotropy, and plasma background flows are taken into account and two different pellet sizes are studied. It has been observed experimentally that injecting pellets during the pedestal build-up does not always lead to ELM triggering~\cite{Lang_2014}. For the present article, we solely concentrate on injections that manage to trigger an ELM. Injections performed at earlier times during the pedestal build-up lie outside the scope of the present study and are investigated separately in Ref.~\cite{Futatani2020}, where the experimentally observed lag-time in the ELM cycle during which ELM triggering is not possible is compared to simulation results.

The first experiments on ELM pacing and mitigation by pellet injection were performed at ASDEX Upgrade (AUG)~\cite{Lang2004_1} demonstrating an increase of the ELM frequency by more than 50\% in a reliable manner. With high pellet injection frequencies, undesired fuelling effects were observed due to the comparably large pellet sizes available in this machine. Pellet injection was shown to trigger an ELM when the pellet-induced seed perturbation was roughy half-way between the separatrix and the pedestal top ($\sim3~\mathrm{cm}$ measured along the pellet trajectory), or ${\sim0.1~\mathrm{ms}}$ after the pellet crossed the separatrix (for injection velocities of ${\sim600~\mathrm{m/s}}$)~\cite{Kocsis_2007}. Key findings from AUG with carbon wall (AUG-C) were confirmed at larger machines like DIII-D~\cite{Baylor2013PRL} and JET~\cite{Romanelli2009}. Investigation of pellet-triggered MHD events in AUG-C and JET showed that triggered and spontaneous ELMs display essentially the same features. In particular, the triggered ELM correlates to the spontaneous ELM type occurring at the same plasma conditions~\cite{Lang2008}. 


Experiments with metal walls in JET with ITER-like wall (JET-ILW) and in ASDEX Upgrade with tungsten coated walls (AUG-W) showed that injecting pellets at different times during the ELM cycle are not always able to trigger ELMs~\cite{Lang_2013,Lang_2014,Lang2015}. Said lag-time represents an important uncertainty in terms of pellet ELM pacing in future machines with metal walls, like ITER. Another uncertainty that prevails for the feasibility of pellet injection as an ELM control method is the wetted area of the heat flux deposition on the divertor targets. In particular, from current machines it is known that the wetted area for pellet-triggered ELMs is smaller than that for spontaneously occurring ELMs~\cite{Wenninger2011}. Therefore, even if pellet-triggered ELMs cause smaller energy losses than spontaneous ELMs, the fact that said energy is deposited over a smaller area could mean that there is no reduction in the energy fluency for pellet-triggered ELMs.

The article is structured as follows. Section~\ref{jorek-and-older} provides essential information regarding the physics model and the pellet module in JOREK and it provides a brief overview of previous pellet-triggered ELM simulations produced with JOREK. Section~\ref{setup} describes the simulation set-up and the different pellet injection parameters used. A thorough description of the temporal dynamics of spontaneous and triggered ELMs, including detailed analysis of the observed similarities and differences, is provided in section~\ref{elms}. Section~\ref{compare} is focused on comparing several spontaneous and pellet-triggered ELMs in terms of ELM sizes, toroidal mode spectra, divertor incident power, peak heat fluxes, and energy fluency. The article ends with the main conclusions and outlook to future work. 

\section{JOREK and overview of existing simulations}\label{jorek-and-older}

The JOREK code is a thoroughly tested MHD code (see Section~4 of Ref.~\cite{Hoelzl2020A} for an overview) that is used to non-linearly solve the reduced and full MHD equations. The reduced  MHD model with various extensions~\cite{Orain2013,Pamela2017} has been used for this work. It evolves the coupled equations for poloidal magnetic flux $\psi$, single fluid mass density $\rho$ and temperature  ${T(=T_e+T_i)}$, parallel plasma velocity $v_\parallel$, and electrostatic potential $\Phi$. Equations for the toroidal current density $j_\phi$ and the vorticity $\omega$ are also solved for numerical reasons. Further details of the model are outlined below.

\subsection{JOREK reduced MHD model}\label{jorek}

The reduced MHD model solved with the JOREK code relies on two simplifying assumptions on the magnetic field and on the plasma velocity. The latter is the zeroth order assumption for the plasma velocity perpendicular to the magnetic field ${\bm{v_\bot} = \bm{v}_\mathrm{E\times B}}$, and allows a potential formulation for $\bm{v_\bot}$ through the electrostatic scalar potential, i.e. ${\bm E = - \grad \Phi}$. The former is the consideration of a static toroidal magnetic field that only varies with major radius described by ${B_\mathrm{tor}(R) = B_\mathrm{axis} R_\mathrm{axis} / R = F_0 / R}$. This assumption simultaneously eliminates one dynamic variable and the fast magnetosonic wave from the system. 

Introducing the assumptions above into the visco-resistive MHD equations, and considering diffusive particle and heat transport, results into the base reduced MHD equations in JOREK~\cite{Huysmans2007}. The perpendicular diffusive particle and heat transport used in JOREK are ad-hoc diffusion profiles meant to represent anomalous transport. The parallel heat diffusion is temperature dependent and it is set by the realistic Spitzer-H\"arm coefficients.  It is possible to include the two-fluid diamagnetic drift effect onto the base reduced MHD equations. In order to do so, the representation for the perpendicular plasma velocity is changed to $\bm v_\bot = \bm v_\mathrm{ExB} + \bm v^*_\mathrm{ion}$. Due to this extension to reduced MHD, the radial electric field well ubiquitous to the edge of H-mode plasmas (and roughly proportional to the ion diamagnetic drift velocity $v^*_\mathrm{ion}\sim\nabla p_i/n$) can be considered in JOREK simulations~\cite{Groebner_Burrell_Seraydarian_1990,cavedon2017pedestal}. The presence of this radial electric field has important consequences regarding the stability of the underlying instabilities that give rise to ELMs~\cite{rogers1999diamagnetic}. These extensions allowing to incorporate realistic plasma flows are used in the present work.

The reduced MHD model is further extended to include the self-generated neoclassical bootstrap current that is formed due to collisions between trapped and passing particles. The bootstrap current density increases with increasing pressure gradient. To include this neoclassical effect in JOREK, a source term determined by the Sauter formula of the bootstrap current is considered in the induction equation~\cite{Sauter1999,Sauter2002}. The resulting equations may be found in~\cite{Pamela2017}.

\subsection{Pellet module}\label{ngs}

In order to study the influence of pellet injection onto the plasma, JOREK features a so-called pellet module. The pellet module represents the pellet particles that will be deposited to the bulk plasma as a localised adiabatic 3D density source. The density source is poloidally localised to a narrow area, and it is stretched to span a user-defined toroidal arc. The adiabatic density source moves with the pellet position, and its time-dependent amplitude results from the neutral gas shielding ablation model as described in~\cite{Futatani2014}. For a given time point this model describes the number of particles that are ionised and become part of the bulk plasma. 

The pellet travels in a straight line following the direction of the predetermined injection velocity. The adiabatic 3D density source locally increases the plasma density and, in turn, directly causes a reduction in temperature. In the pellet ablation cloud (that stretches along the magnetic field lines), the pressure increases as the temperature is partially restored by fast parallel electron heat conduction (${\tau_{\chi_\parallel}= (2\pi R q)^2/\chi_\parallel \sim 0.1~\mathrm{\mu s}}$) while the density redistribution within a flux surface happens on the longer time scale of parallel convection with the ion sound speed (${\tau_s= 2\pi R q/c_s \sim 1~\mathrm{ms}}$). 

\subsection{Relation to previous simulation work}

Edge localized mode physics was studied in various ways already using the JOREK code – first in Ref.~\cite{Huysmans2007}. Further relevant simulations with JOREK include: spontaneous ELMs with realistic plasma background flows~\cite{Orain2013}, RMP penetration~\cite{Nardon2007,Orain2017}, investigation of ELM-RMP interactions~\cite{Becoulet2014}, Quiescent H-Mode~\cite{LiuF2015}, triggering of ELMs by vertical magnetic kicks~\cite{Artola2018}, and a direct comparison of the divertor heat fluence caused by spontaneous ELMs to experimental scaling laws~\cite{Pamela2017}. Pellet ELM triggering has also been studied with JOREK before, providing an explanation for the mechanism of pellet ELM triggering by a localised increase of the pressure in the re-heated ablation cloud and including experimental comparisons to JET and DIII-D~\cite{Huysmans2009,Futatani2014,Futatani2019}. An overview of ELM related non-linear MHD simulations worldwide is given in Ref.~\cite{Huijsmans2015}. However, the field has evolved rapidly since the publication of that article. A recent overview of ELM and ELM control simulations with JOREK is given in Section~5 of Ref.~\cite{Hoelzl2020A}.

For ASDEX Upgrade, in particular the localised structures forming during ELM crashes~\cite{Hoelzl2012A}, non-linear mode coupling associated with an ELM crash~\cite{Krebs2013}, the toroidal structure of an ELM crash~\cite{Hoelzl2018}, and ELM control via resonant magnetic perturbation fields~\cite{Orain2019} have been studied. Recently, type-I ELM cycles and the triggering mechanism responsible for the violent onset of the ELM crash were studied for the first time~\cite{Cathey2020}. In the present article,  pellet-triggered ELM simulations are compared to spontaneous ELMs from the aforementioned article and new dedicated simulations. 

The present article goes beyond previous studies of pellet-triggered ELMs by producing a direct comparison between state-of-the-art simulations of type-I ELM cycles and pellet-triggered ELMs where both are performed using the same extended MHD model including ExB and diamagnetic background flows and the same plasma conditions with realistic plasma parameters.

\section{Simulation set-up and parameters}\label{setup}

Recent simulations of type-I ELM cycles show that the seed perturbations out of which instabilities grow prior to the ELM have an important effect on the dynamics of the non-linear dynamics and properties of the crash. Namely, starting from arbitrary seed perturbations, instead of self-consistent perturbations, causes larger ELMs when the pedestal build-up is considered. Simulating type-I ELM cycles circumvents this problem since the seed perturbations for all ELMs (except the first) retain the ``memory'' of the previous existence of an ELM, i.e., have non-negligible amplitudes and peeling-ballooning mode structure~\cite{Cathey2020}. The present study uses the first three type-I ELMs from~\cite{Cathey2020} to compare the dynamics of spontaneous ELMs and pellet-triggered ELMs. Additionally, new simulations of type-I ELM cycles with increased toroidal resolution are performed to compare against the pellet-triggered ELMs. 

The spontaneous ELM simulations are set up in the following way. An ideal MHD stable post-ELM equilibrium reconstruction for AUG shot \#33616 is used as initial conditions for the simulations. 
We set stationary heat and particle diffusion (perpendicular to $\bm B$) profiles with a well in the pedestal region to model the H-mode edge transport barrier as well as stationary heat and particle sources such that the pedestal builds up with time and crosses the peeling-ballooning stability boundary. This simplified pedestal build-up does not take into account the dynamical response of anomalous transport, nor does it include the physics of neutrals, which are key for a realistic consideration of the time-evolving particle source profiles. Since we use a single fluid model, the disparate evolution timescales of the electron and ion temperatures is neglected. Plasma resistivity and parallel heat transport are modelled with fully realistic parameters, and with Spitzer and Spitzer-H\"arm temperature dependencies, respectively. Simulations are performed with a high resolution finite element grid in the poloidal direction and the convergence of the results has been verified. 

\begin{figure}
\centering
  \includegraphics[width=0.49\textwidth]{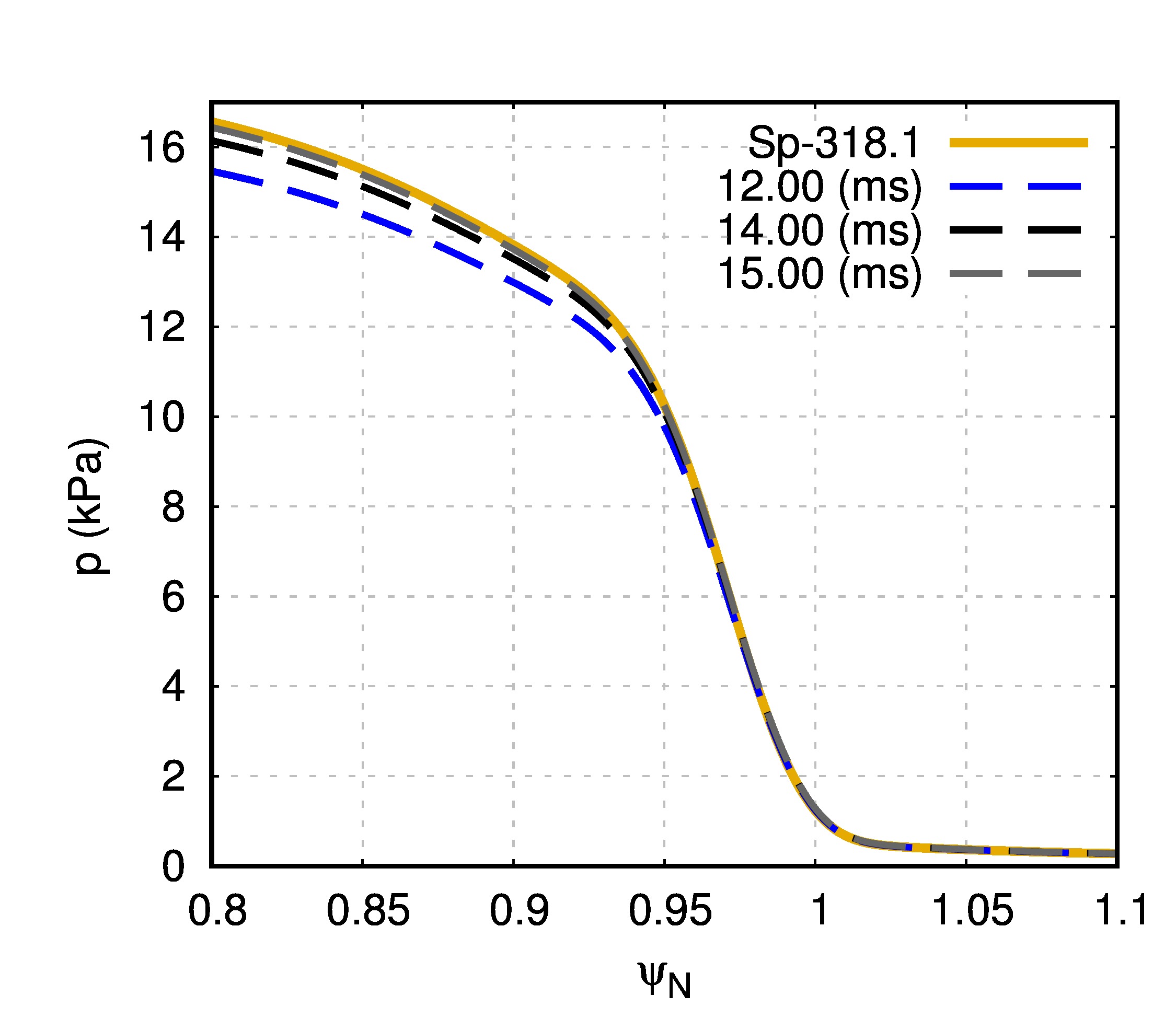} 
\caption{Pressure profiles at the outboard midplane in the pre-ELM stage of a spontaneous ELM (full lines) and at the different pellet injection times (dashed lines).}
\label{fig:pressure}
\end{figure}

With the density and temperature profiles steepening as the simulation evolves, the radial electric field and the neoclassical bootstrap current evolve accordingly. Four pressure profiles at the outer midplane are shown in fig.~\ref{fig:pressure}. These correspond to the pre-ELM state of one spontaneous ELM (full line labelled \texttt{Sp318.1}-- the naming conventions are described later), and to the pellet injection times: $12,\,14,\,\mathrm{and}\,15~\mathrm{ms}$ (dashed lines in blue, black, and gray respectively). 

\subsection{Pellet injection}\label{pellet}

\begin{figure*}
\centering
  \includegraphics[width=0.9\textwidth]{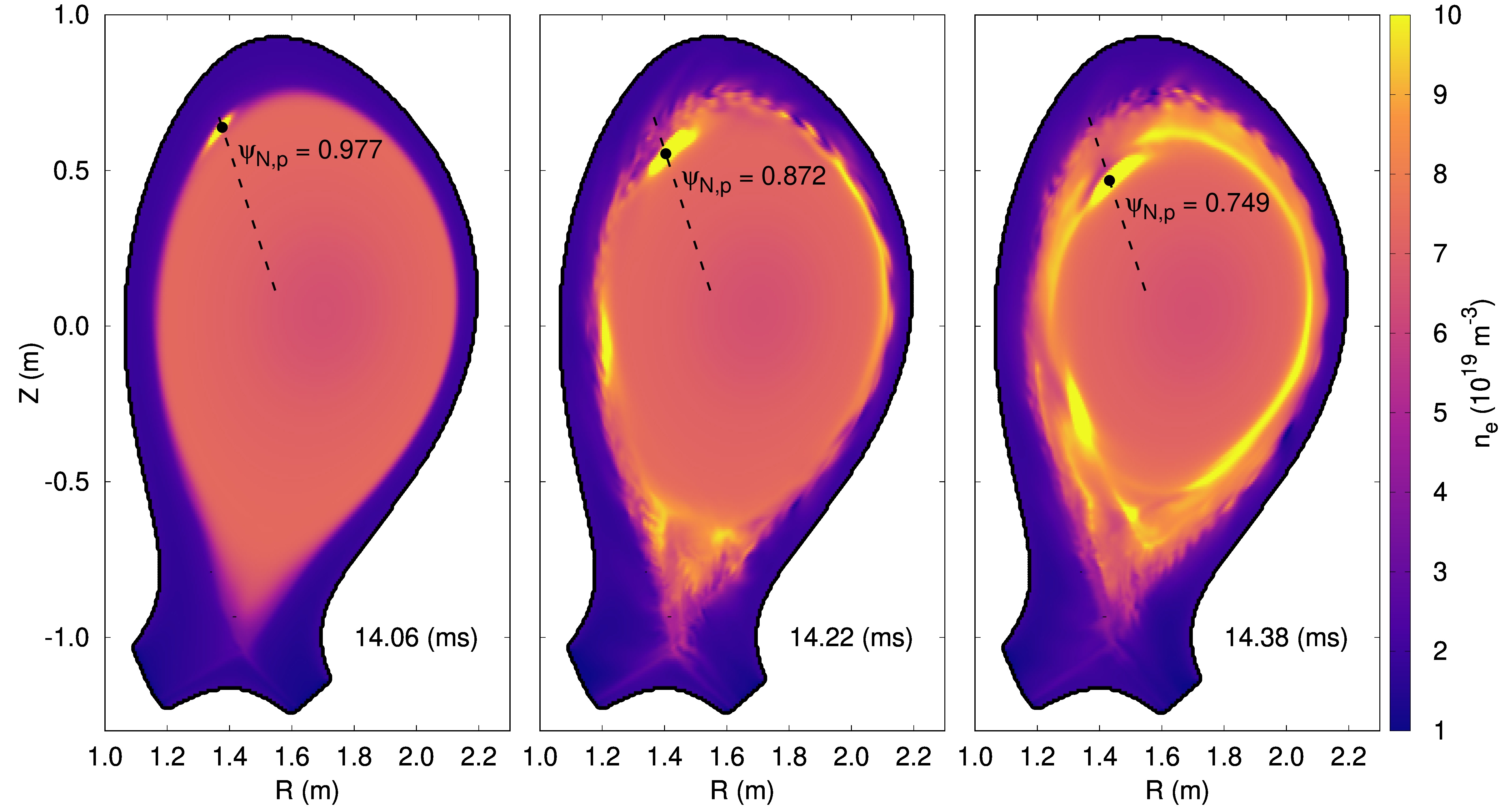} 
\caption{Three different time points of the poloidal profile of the density for \texttt{Tr-15-14ms} -- large ($1.5\times10^{20}\mathrm{D}$ atoms) pellet injection at $14\,\mathrm{ms}$. The redistribution of the excess density introduced by the pellet takes place due to parallel convection at the local sound speed. The pellet trajectory is shown as black dashed line. The pellet location at the three different time points is ${\psin=0.977}$, $0.872$, and $0.749$ respectively.}
\label{fig:pellet}
\end{figure*}

The AUG pellet injection system has an injection angle of ${\alpha_\mathrm{injection}=72^\circ}$, and it can handle pellets of different nominal sizes and velocities. These correspond to ${1.5\times10^{20}, 2.4\times10^{20}}$, and ${3.7\times10^{20}\mathrm{D}}$ atoms per pellet and ${v_\mathrm{p}=240, 560, 900}$, and $1040~\mathrm{m/s}$. However, material is lost on the guide tube as the pellet travels to the plasma and, ultimately, the pellet sizes arriving at the plasma show significant variations from their nominal sizes. These variations are most pronounced with increasing pellet injection velocity and, on average, account for roughly $50\%$ of the nominal mass lost for ${v_\mathrm{p}=560~\mathrm{m/s}}$~\cite{Lang2003}. For the pellets used in the present simulations, we chose ${v_\mathrm{p}=560~\mathrm{m/s}}$ and two different pellet sizes of ${0.8\times10^{20}}$ and ${1.5\times10^{20}\mathrm{D}}$ atoms per pellet (``small'' and ``large''). These values reflect particle content \textit{after losses} in the guide tube and therefore correspond approximately to the range of accessible pellet sizes in the experiment.

The pellet module describes the pellet moving straight with constant velocity. The simplifying assumption of a constant pellet velocity is motivated by AUG experimental observations~\cite{Kocsis_2007}. Using the same initial position and injection velocity and angle, the two different pellet sizes are injected at $12$ and $14~\mathrm{ms}$. The small pellet is additionally injected at $15~\mathrm{ms}$. We adopt a naming convention for the pellet-triggered ELMs using the pellet size and injection time. For example, to refer to the ELM triggered by injecting the small ($0.8\times10^{20}\mathrm{D}$ atoms) pellet at $14~\mathrm{ms}$, we use the name \texttt{Tr-08-14ms}. Accordingly, the large pellet injected at $12~\mathrm{ms}$ is named \texttt{Tr-15-12ms}.

Figure~\ref{fig:pellet} shows poloidal planes (at the toroidal angle corresponding to the pellet injection) of the electron number density $n_e$ at three different time points after a large pellet has been injected at $14~\mathrm{ms}$, i.e. \texttt{Tr-15-14ms}. The pellet trajectory is also plotted with a dashed black line. The first frame shows the density $60~\mathrm{\mu s}$ after the pellet was injected. At this time, no observable non-axisymmetric perturbations (other than the pellet) are seen, and pellet is located inside of the separatrix in the radial position $\psi_\mathrm{N}=0.977$. Shortly thereafter non-axisymmetric MHD activity is prompted and the ELM is triggered. The time of ELM onset is $14.09~\mathrm{ms}$ and the pellet position at the ELM onset is $\psin=0.958$. The plot of $n_e$ during the pellet-triggered ELM crash can be observed in the second and third frames of fig.~\ref{fig:pellet} at $14.22~\mathrm{ms}$ and $14.38~\mathrm{ms}$, respectively. In these frames, strong ($\delta n_e/n_e \sim 1$) peeling-ballooning modes at the edge of the confined region are observed. 

\subsection{Toroidal mode numbers} \label{mode-numbers}

For the spontaneous ELMs simulation of ~\cite{Cathey2020}, which required resolving the fast timescales related to the ELMs (${\tau_\mathrm{A}\sim0.5~\mathrm{\mu s}}$) and the slow time-scales of the pedestal build-up (${\tau_\mathrm{ped}\sim 1/f_{\mathrm{ELM}}\sim10~\mathrm{ms}}$), the even toroidal mode numbers until ${\mathrm{n}=12}$ were considered, i.e. the toroidal periodicity chosen for those simulations was ${\mathrm{n_{period}}=2}$. This is equivalent to simulating only half of the torus, and it was chosen in order to reduce the computational cost of the simulations -- while maintaining the cyclical dynamics. Recent code optimizations~\cite{Holod2020} allow to increase the toroidal resolution for the present work. Thus, dedicated spontaneous ELM simulations were performed for this article with higher maximum toroidal mode numbers, $\mathrm{n_{max}}=15$, $18$, and $20$. For these dedicated simulations the ELM crashes become faster and more violent. For this reason, we focus the comparison on these new spontaneous ELM crash simulations with increased $\mathrm{n_{max}}$. Where comparisons are based on simulations from Ref.~\cite{Cathey2020}, we restrict them to time-integrated quantities as they are less dependent on toroidal resolution.

The new spontaneous ELM simulations with increased $\mathrm{n_{max}}$ include the toroidal mode numbers ${\mathrm{n}=0-3-15}$, ${\mathrm{n}=0-3-18}$ and ${\mathrm{n}=0-2-20}$. The ELM crashes used for the comparisons in this article are labelled reflecting this toroidal resolution: The $\mathrm{j^{th}}$ spontaneous ELM simulated with the following toroidal mode numbers ${\mathrm{n}=0-\mathrm{n_{period}}-\mathrm{n_{max}}}$ is named \texttt{Sp-}$\mathrm{n_{period}n_{max}}$\texttt{.j}. As an example, the first spontaneous ELM simulated with the toroidal mode numbers ${\mathrm{n}=0-3-18}$ is named \texttt{Sp-318.1}. Finally, a simulation with a further increased resolution including the toroidal mode numbers ${\mathrm{n}=0-3-30}$ was performed to check convergence. The relative difference between the ELM-related thermal energy lost between the ELM crash simulated with ${\mathrm{n}=0-3-15}$ and ${\mathrm{n}=0-3-30}$ was found to be $\sim4~\%$.

When simulating pellet injection, it is not possible to use a ``periodicity'' greater than $1$ because the $\mathrm{n=1}$ toroidal mode number is always needed and typically observed to be dominant~\cite{Futatani2014,Futatani2019}. As a result, the toroidal discretisation in JOREK must contain the entire mode spectrum, i.e. the entire torus has to be simulated. The toroidal mode numbers included for all pellet-triggered ELM simulations presented in this work are ${\mathrm{n}=0-1-12}$. As can be seen from the energy spectra shown in Ref.~\cite{Futatani2020}, the ${\mathrm{n}=12}$ mode is strongly sub-dominant in all cases providing a justification for this choice. Higher toroidal resolutions will only be affordable after future code optimizations.


\section{Spontaneous and pellet-triggered ELMs}\label{elms}

This section is divided into four parts. In subsection~\ref{spontaneous}, we present a detailed description of the temporal dynamics of a representative spontaneous ELM crash (\texttt{Sp-318.1}), and in subsection~\ref{triggered} of a representative pellet-triggered ELM (\texttt{Tr-08-14ms}). Each subsection details the evolution of the energies of the non-axisymmetric perturbations present in the simulations, the power that is incident on the simplified divertor targets, the time evolving reconnecting magnetic field, and other dynamical quantities of interest. Thereafter, in subsection~\ref{sidebyside} a side-by-side comparison in terms of the heat fluxes between the representative pellet-triggered and spontaneous ELMs is presented. Finally, subsection~\ref{modestructures} describes the differences and similarities of the non-axisymmetric mode activity present during the ELM crash between \texttt{Sp-318.1} and \texttt{Tr-08-14ms}. A systematic comparison of all cases is described later in section~\ref{compare}.

\subsection{Spontaneous ELMs}\label{spontaneous}

\begin{figure}
\centering
  \includegraphics[width=0.49\textwidth]{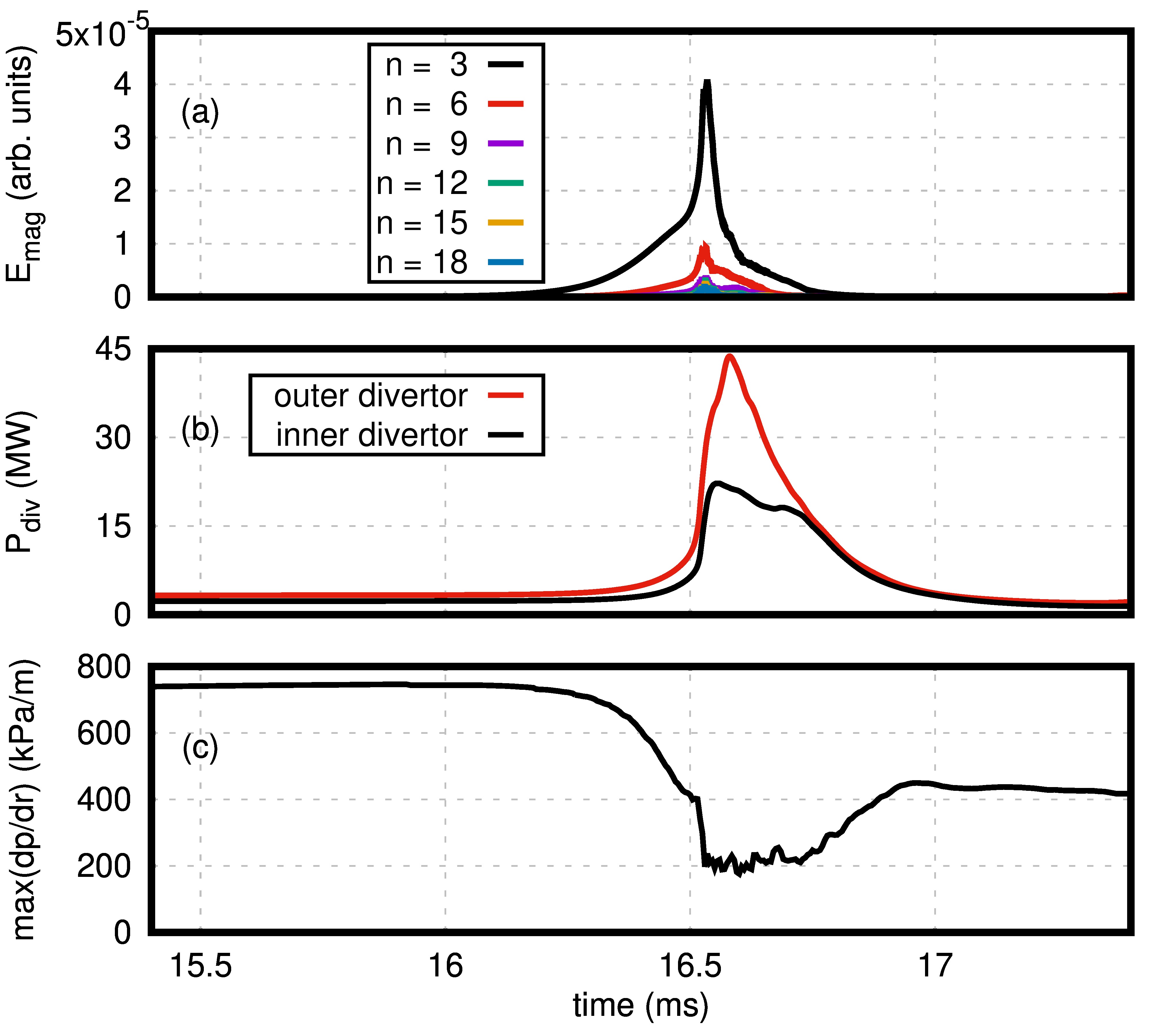} 
\caption{Time evolution of the magnetic energies (a), incident power on the inner and outer divertors (b), and change in the toroidally averaged outboard midplane maximum edge pressure gradient (c) during a spontaneous ELM.}
\label{fig:spontaneousELM}
\end{figure}

Here, we provide a detailed look at the dynamics of a representative spontaneous ELM crash. In this section, we focus on the first type-I ELM from the simulation with $\mathrm{n}=0-3-18$. This ELM crash takes place at $16~\mathrm{ms}$, and the pre-ELM pressure profile was shown in fig.~\ref{fig:pressure}. The magnetic energy of the non-axisymmetric perturbations, the incident power on the inner and outer divertors, and the toroidally averaged outboard midplane maximum edge pressure gradient are shown in fig.~\ref{fig:spontaneousELM}(a-c). During the ELM crash the thermal energy stored inside the separatrix is reduced from $421~\mathrm{kJ}$ to $388~\mathrm{kJ}$, i.e. the ELM size is ${\Delta W_\mathrm{ELM}=33~\mathrm{kJ}}$ and the relative ELM size is ${\Delta E_\mathrm{ELM}=\Delta W_\mathrm{ELM}/W_\mathrm{preELM}\approx7.8\%}$. As a result of the energy lost from the confined region, the incident power to the inner and outer divertors increases sharply from the inter-ELM values of $2.2$ and $3.3~\mathrm{MW}$ to peak values of $22.2$ and $43.8~\mathrm{MW}$, respectively. After the crash, $P_\mathrm{div,in/out}$ return to the inter-ELM values within roughly $0.6~\mathrm{ms}$.

The heat flux that arrives at the divertor targets is ${q(t,s,\phi)}$, where $t$ is time, $s$ is the distance along the target, and $\phi$ is the toroidal angle. The total power that reaches a given divertor target is defined by ${P_\mathrm{div}=\int_0^{2\pi}\int_{s_0}^{s_\mathrm{max}}q(t,s,\phi) R\,ds\,d\phi}$, where $R$ is the major radius. During the peak of the ELM crash, the power that reaches the outer divertor target $P_\mathrm{div,out}$ is roughly twice the power reaching the inner target $P_\mathrm{div,inn}$. This power asymmetry is observed for all the simulated spontaneous and pellet-triggered ELMs. In experiments under similar conditions, i.e., with the ion ${\bm B \times \grad B}$ direction pointing to the active X-point, the inner divertor receives more energy than the outer divertor~\cite{Eich2005}. Without a proper inclusion of ExB and diamagnetic background flows, this discrepancy would be more pronounced~\cite{Orain2015}. The main reason for the remaining differences lies in the simplified model for the SOL used here. Separate efforts are underway to address this issue, but are not part of the present work.

\begin{figure}
\centering
  \includegraphics[width=0.49\textwidth]{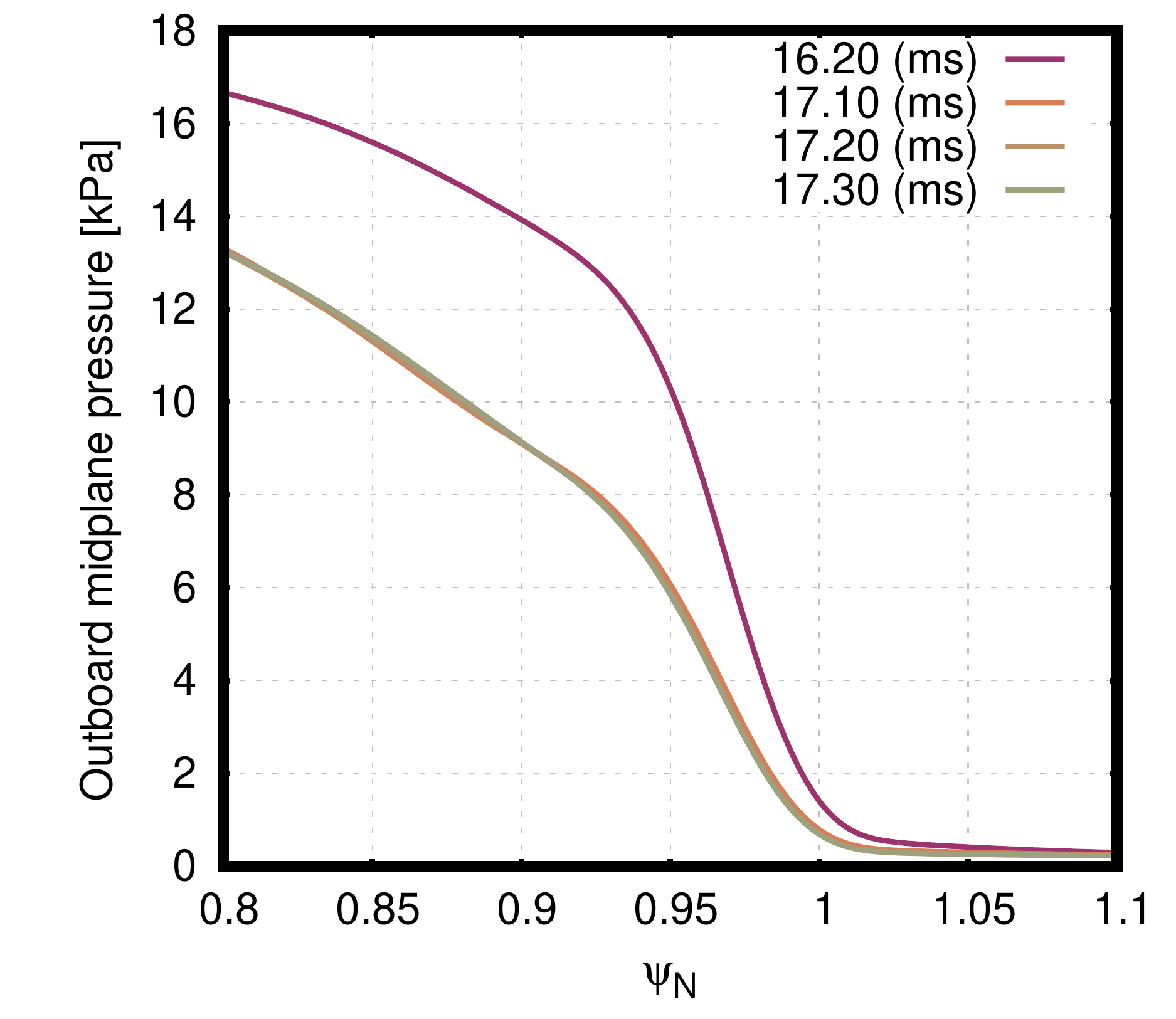} 
\caption{Outer midplane pressure profiles before and after a spontaneous ELM crash (\texttt{Sp-318.1}). The pressure at the radial location $\psi_N=0.95$ goes from $\sim10.3~\mathrm{kPa}$ to $\sim5.9~\mathrm{kPa}$.}
\label{fig:spontaneousELM-ped}
\end{figure}

The maximum (toroidally averaged at the outer midplane) pressure gradient crashes rapidly as a result of the ELM as shown in fig.~\ref{fig:spontaneousELM}(c). This reduction causes the drive for the underlying instabilities to be removed and for the ELM crash to conclude shortly thereafter. This allows the maximum $\grad p$ to recover quickly until it transiently stagnates at ${\sim16.8~\mathrm{ms}}$ due to the excitation of post-cursor modes, which are not the subject of the present work. Afterwards, the pedestal builds up further until the next ELM appears (not shown). The pedestal profiles before (${16.20~\mathrm{ms}}$) and after ($17.10,\,17.20,\,\mathrm{and}\,17.30~\mathrm{ms}$) the ELM crash are shown in fig.~\ref{fig:spontaneousELM-ped}, where it is evident that the spontaneous ELM is significantly reducing the edge pedestal pressure. Together with the depletion of the pressure pedestal, the radial electric field and the bootstrap current density are similarly reduced.

\begin{figure}[!ht]
\centering
  \includegraphics[width=0.49\textwidth]{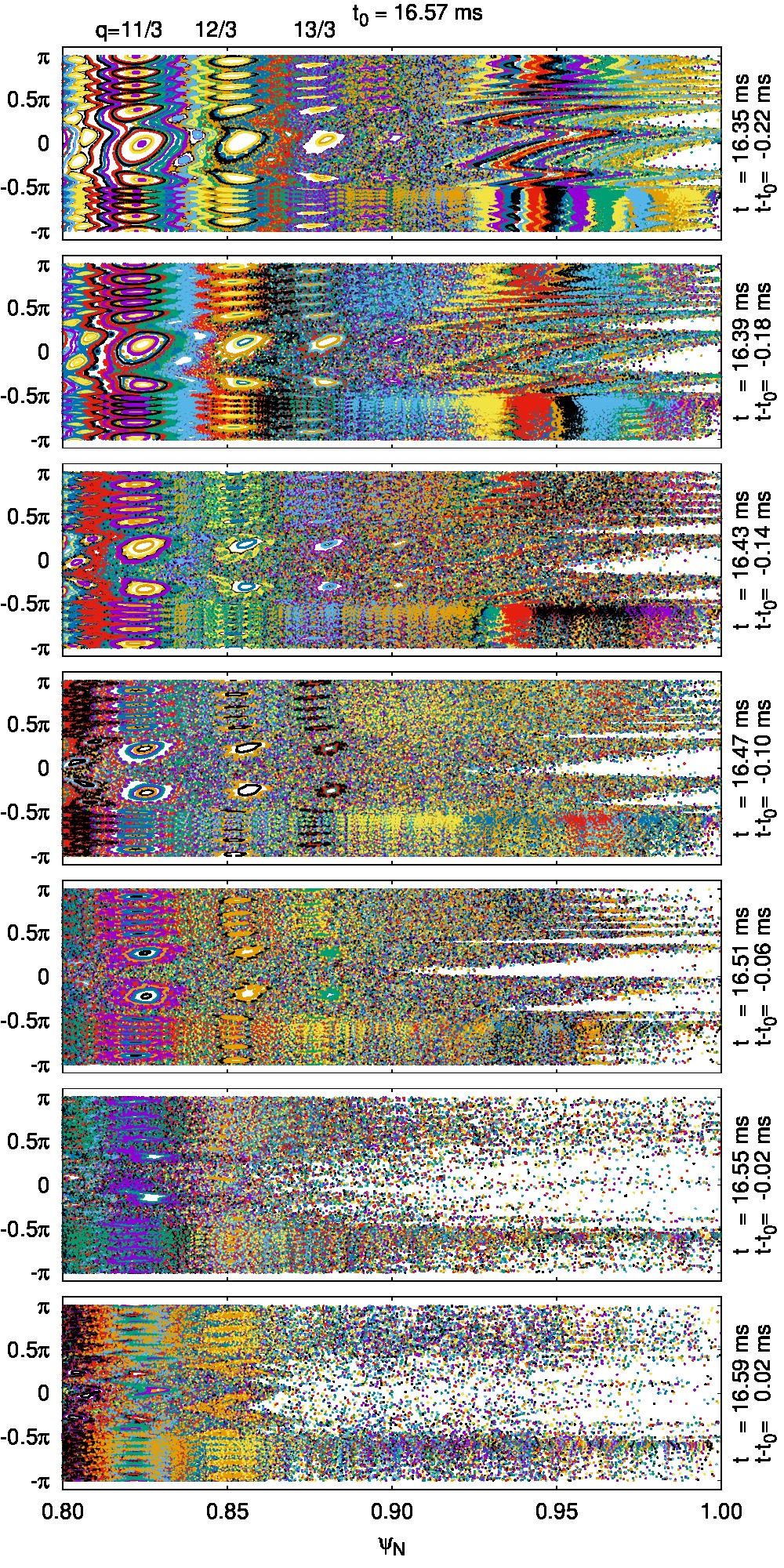} 
\caption{Seven Poincar\'e plots at different times with respect to ${t_0=16.57~\mathrm{ms}}$. The chosen times are separated by ${0.04~\mathrm{ms}}$. Precursor modes cause the non-axisymmetric topology observed in the earliest plots (top panels). The growing perturbations cause the region with short connection length (white region) to penetrate further inward until ${\psin\approx0.87}$ (bottom panels).}
\label{fig:spontaneousELM-poincare}
\end{figure}

Poincar\'e plots for this spontaneous ELM are shown in fig.~\ref{fig:spontaneousELM-poincare} at different times for the radial region from ${\psin=[0.8-1.0]}$. Each Poincar\'e plot is produced by following $400$ magnetic field lines with different radial and poloidal initial positions for $3000$ toroidal turns. The positions of the field lines when crossing the toroidal position $\phi=0$ are plotted. Different colors are used for each field line such that the ``mixing of colors'' visualizes the radial diffusion of field lines during stochastisation. The y-axis corresponds to the geometrical poloidal angle $\theta$ with the outboard midplane located at $\theta=0$. The time of maximum heat flux $\Podiv$ to the outer divertor is ${t_0 = 16.57~\mathrm{ms}}$, and the different times correspond to ${t-t_0 = [-0.22:0.04:0.02]~\mathrm{ms}}$.

At the earliest time plotted, ${t-t_0=-0.22~\mathrm{ms}}$, the energy of the non-axisymmetric perturbations is already large, as can be seen in fig.~\ref{fig:spontaneousELM}(a), and the density perturbations are comparable to the background plasma ($\delta n/n\sim 1$). These perturbations are peeling-ballooning modes which are responsible for the non-axisymmetric magnetic fields observed in fig.~\ref{fig:spontaneousELM-poincare}. A strong deformation of the flux surfaces at the very edge of the plasma (${\psin \gtrsim 0.88}$) causes an increase in stochastic cross-field transport through parallel heat (Spitzer-H\"arm) diffusion that may be evidenced by the increases to the divertor incident power shown in fig.~\ref{fig:spontaneousELM}(b). In the region ${\psin\approx[0.80-0.88]}$ three sets of magnetic islands located at the ${q=11/3,\,12/3,\,\mathrm{and}\,13/3}$ rational surfaces are present at ${t-t_0=-0.22~\mathrm{ms}}$. The amplitude of the peeling-ballooning modes increases further thereby causing the stochastic magnetic topology to erode further inwards. The increased stochastisation around the time of maximum divertor heat flux (the two bottom plots of fig.~\ref{fig:spontaneousELM-poincare}) is clearly visible. Field lines in the outer plasma regions $\psin\gtrsim0.87$ reach the divertor targets with a very short connection length in this phase, as reflected by the strongly reduced number of points visible close to the plasma edge.

\begin{figure}
\centering
  \includegraphics[width=0.49\textwidth]{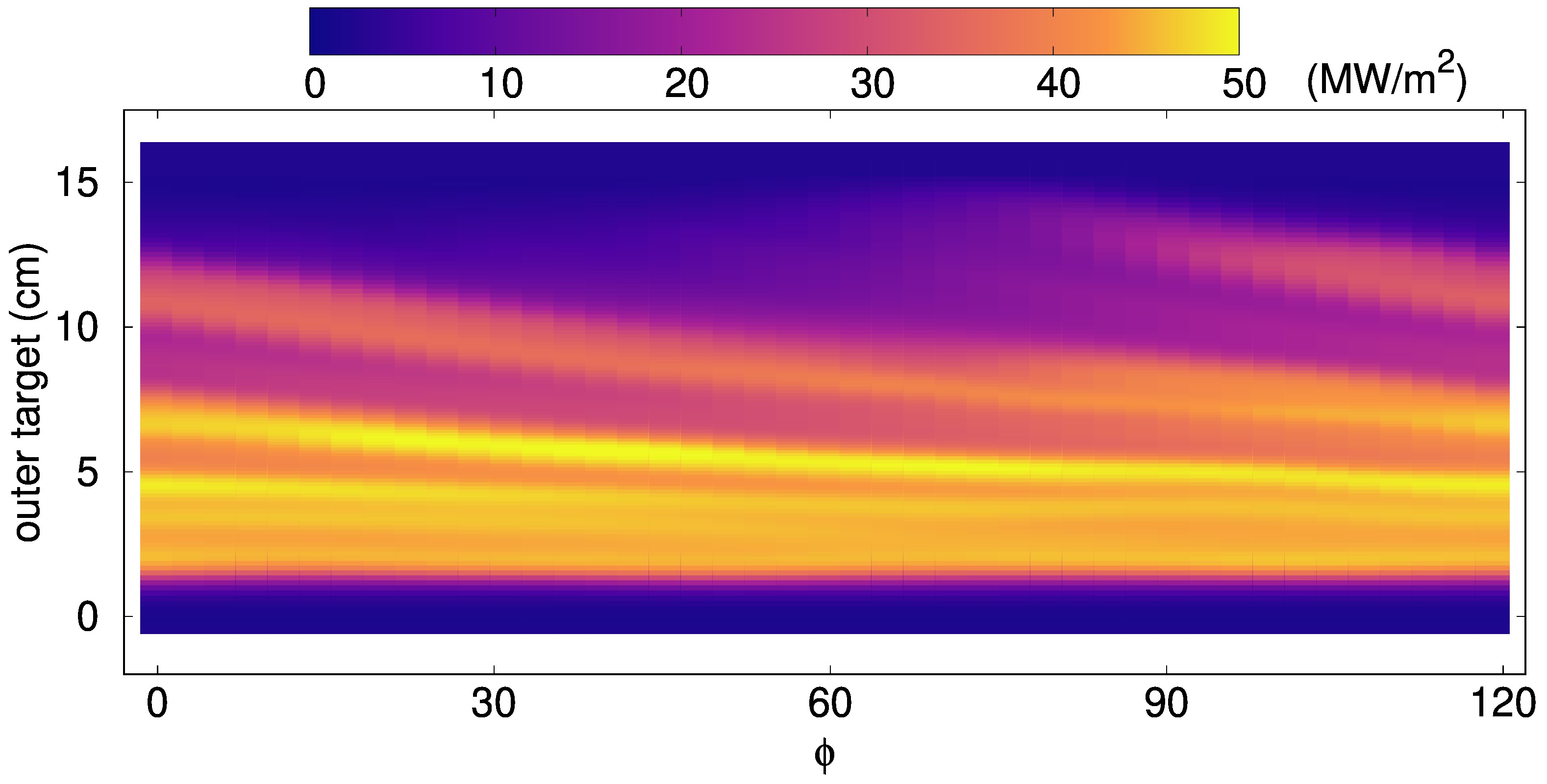} 
\caption{Outer divertor heat flux profile along different toroidal angles at the time of maximum $\Podiv$, $t_0=16.57~\mathrm{ms}$. The footprint corresponding to several strike lines can be observed. The wetted area at this time point is ${A_\mathrm{wet}\approx0.95~\mathrm{m^2}}$.}
\label{fig:spontaneousELM-heatflux}
\end{figure}

Figure~\ref{fig:spontaneousELM-poincare} features characteristic structures in the outermost edge. These lobe structures represent splitting of the strike lines that hit the divertor targets (and are shown clearly later on in fig.~\ref{fig:compare-poincare}). Such splitting is a common experimental observation, and evaluating quasi-toroidal mode numbers has been accomplished through such observations~\cite{Eich2005,Jakubowski2009}. At the time of maximum incident power on the outer divertor, ${t=t_0}$, for instance, the heat flux impinging on different toroidal angles\footnote{The range of toroidal angles is limited to ${\phi=[0-120]^\circ}$ because this simulation only considers toroidal mode numbers which are multiples of $3$, i.e. one third of the tokamak.} of the outer divertor is shown in fig.~\ref{fig:spontaneousELM-heatflux}, and the strike line splitting can be inferred from the slightly slanted stripes in the heat flux profile.

An important feature in terms of the heat flux onto the divertor target, which is related to the strike line splitting, is the area over which the energy is deposited, i.e. the wetted area
\begin{align}
    A_\mathrm{wet}=\frac{\int_0^{2\pi}\int q(s,t=t_0)dsRd\phi}{\mathrm{max}(q(s,t=t_0))}=\frac{P_\mathrm{div}(t=t_0)}{\mathrm{max}(q(s,t=t_0))}.\nonumber
\end{align}
For the type-I ELM crash discussed here (\texttt{Sp-318.1}), the wetted area of the outer divertor at the time corresponding to fig.~\ref{fig:spontaneousELM-heatflux} is ${A_\mathrm{wet}\approx0.95~\mathrm{m^2}}$.




\subsection{Pellet-triggered ELMs}\label{triggered}

\begin{figure}[!h]
\centering
  \includegraphics[width=0.49\textwidth]{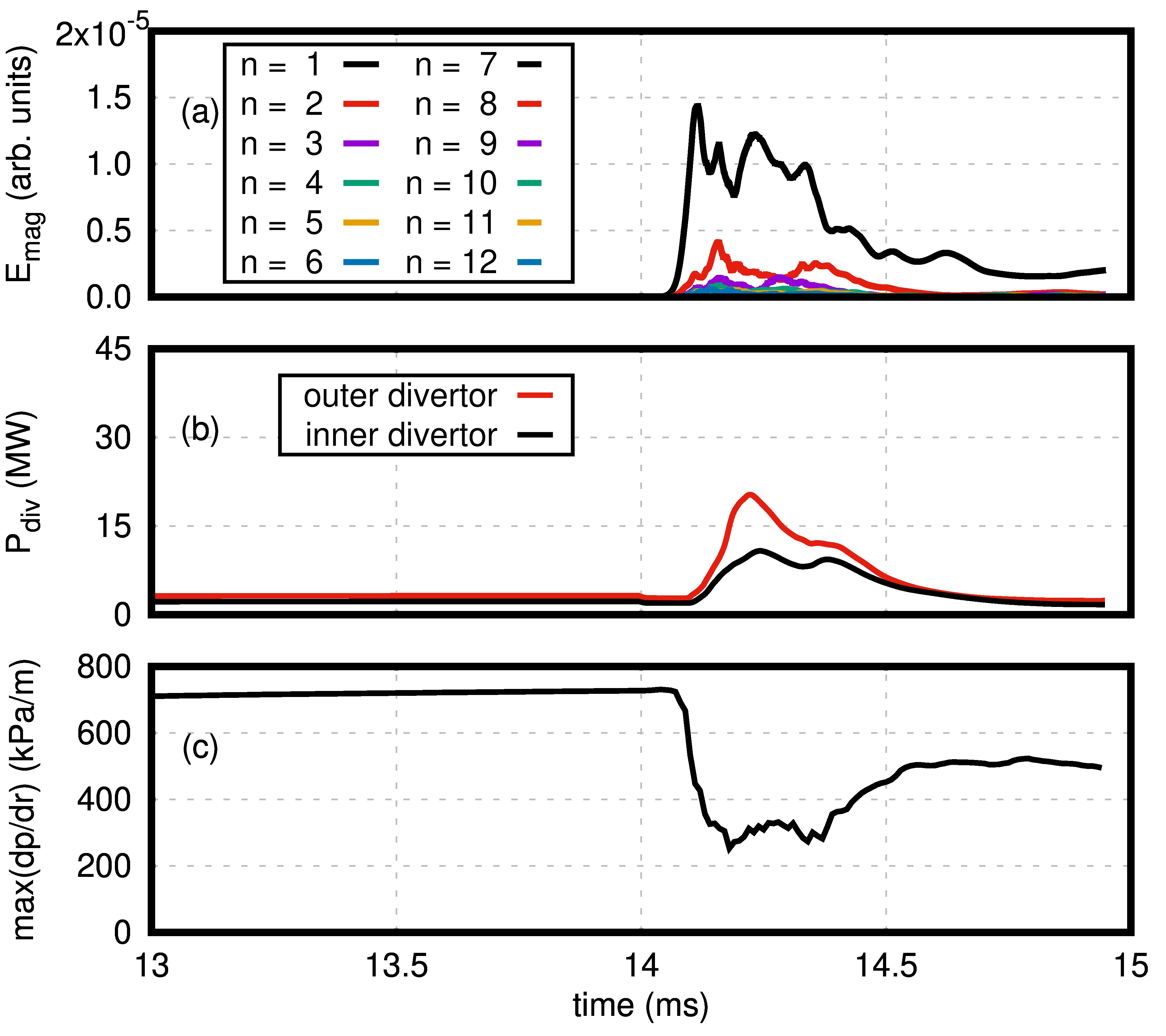} 
\caption{Time evolution of the magnetic energies (a), incident power on the inner and outer divertors (b), and change in the maximum edge pressure gradient (c) during a pellet-triggered ELM corresponding to a pellet of $0.8\times10^{20}~\mathrm{D}$ atoms at $14~\mathrm{ms}$.}
\label{fig:triggeredELM}
\end{figure}

Now, we shift the focus to the temporal dynamics relevant to pellet-triggered ELMs. To do so, we present a detailed description of a representative ELM triggering case where a pellet of $0.8\times10^{20}~\mathrm{D}$ atoms is injected at $14~\mathrm{ms}$ (\texttt{Tr-08-14ms}), i.e. $2~\mathrm{ms}$ before the spontaneous ELM crash would appear. Like all the pellet-triggered ELMs present in this work, the toroidal mode numbers included in the simulation are ${\mathrm{n}=0-1-12}$. At $14~\mathrm{ms}$, the thermal energy stored inside the plasma at the time of injection is ($\sim5~\mathrm{kJ}$) less than before the spontaneous ELM at $16~\mathrm{ms}$ because of the continuous pedestal build-up. In fig.~\ref{fig:triggeredELM}(a-c) we show the normalised magnetic energy of the non-axisymmetric perturbations, the power that is incident onto the inner and outer divertor targets, and the toroidally averaged maximum pressure gradient. 

The ${\mathrm{n}=1}$ magnetic perturbation is dominant as shown in fig.~\ref{fig:triggeredELM}(a). As the pellet is introduced in the simulation, the ${\mathrm{n}=1}$ perturbation begins to grow and it is followed by perturbations characterised by all the remaining toroidal mode numbers. The thermal energy expelled by \texttt{Tr-08-14ms} is roughly $18~\mathrm{kJ}$, which corresponds to a relative ELM size of $4.4\%$. This energy ultimately reaches the divertor targets, as observed in fig.~\ref{fig:triggeredELM}(b). The peak incident power onto the inner and outer divertors caused by this pellet-triggered ELM are $10.8$ and $20.2~\mathrm{MW}$, respectively. 

\begin{figure}
\centering
  \includegraphics[width=0.49\textwidth]{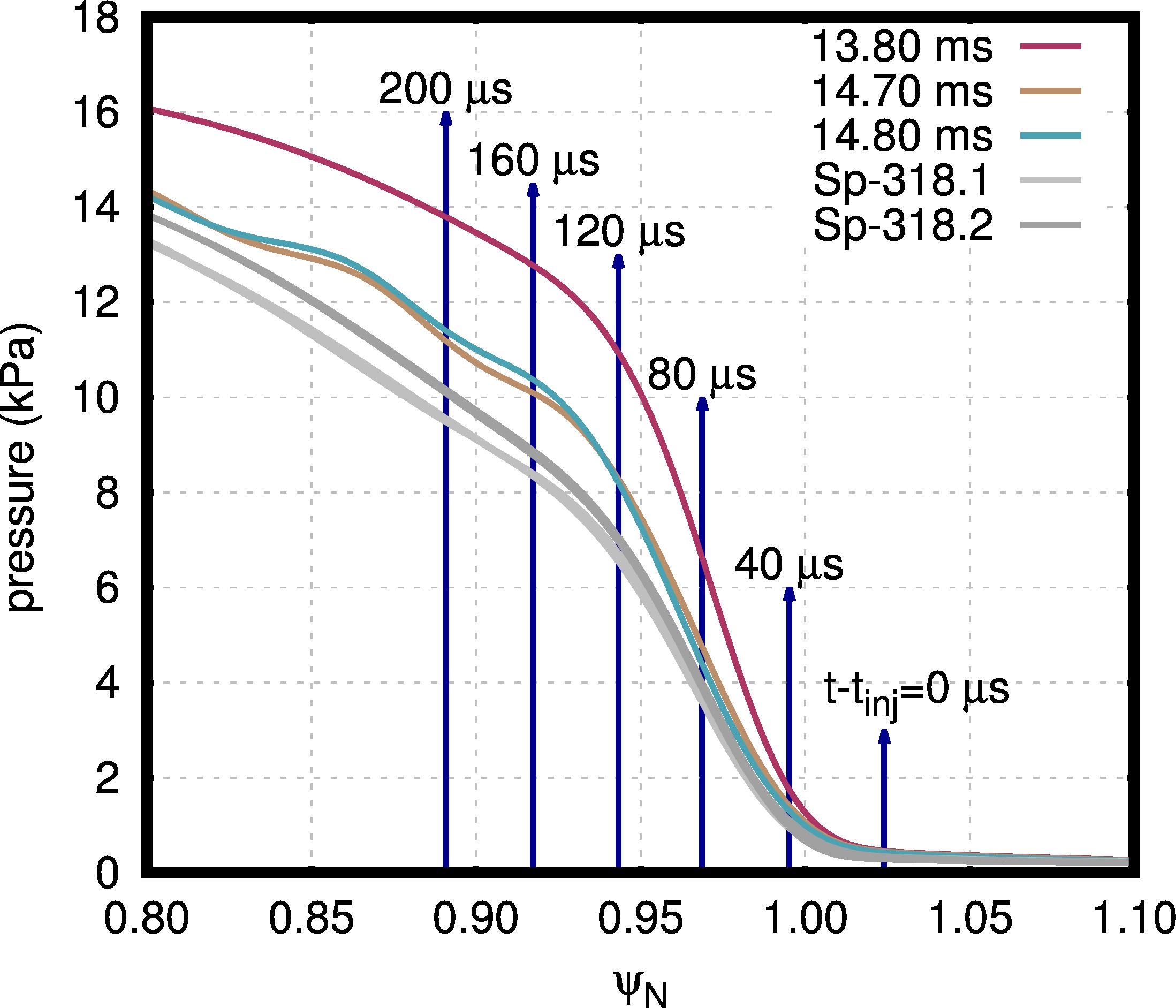} 
\caption{Outer midplane pressure profiles before (in dark red) and after \texttt{Tr-08-14ms} (in yellow and green), and after \texttt{Sp-318.1} and \texttt{Sp-318.2} (in grey and dark grey, respectively). The post-ELM profiles for the spontaneous ELMs have lower pedestals than \texttt{Tr-08-14ms}. The position of the pellet in terms of the normalised poloidal flux is also shown for 6 different points in time.}
\label{fig:triggeredELM-ped}
\end{figure}

The initial position of the pellet is ${(1.365,0.674)~\mathrm{m}}$, which corresponds to an open flux surface just outside the separatrix: ${\psin=1.019}$. The maximum pressure gradient region is at ${\psin=0.972}$, which is roughly ${4.5~\mathrm{cm}}$ along the pellet trajectory. Considering the pellet velocity of ${v_\mathrm{p}=560~\mathrm{m/s}}$, the time required to reach the maximum pressure gradient location is ${80~\mathrm{\mu s}}$. The pre- and post-ELM pressure profiles for \texttt{Tr-08-14ms}, and post-ELM for \texttt{Sp-318.1} and \texttt{Sp-318.2}, at the outer midplane are shown in fig.~\ref{fig:triggeredELM-ped}. The position of the pellet in terms of $\psin$ for 6 different time points is also plotted. It is observed that the post-ELM profiles for the spontaneous ELMs considered here have lower pedestals than the pellet-triggered ELM. 

As the pellet enters the confined region, the growth rates of the high-n perturbations begin to increase rapidly. This increase takes place due to the excitation of ballooning modes in the high-field side (HFS) and low-field side (LFS) regions. In the absence of pellet injection, ballooning modes are localised at the LFS because the magnetic field curvature (which stabilises ballooning modes and always points to the centre of the device) is parallel to the pressure gradient (which destabilises ballooning modes and always points to the magnetic axis) in the LFS, and it is anti-parallel to $\grad p$ in the HFS. However, in the presence of pellet injection, as the pellet enters the confined region there is a significant local increase of the density taking place due to the pellet ablation. With an adiabatic source, the density increase is related to a temperature decrease. The local temperature is rapidly increased again due to the fast electron parallel heat transport (simulations performed with the realistic Spitzer-H\"arm parallel heat diffusion). The excess density is also transported along the magnetic field lines, but on the much slower time scale of the ion sound speed, as described in section~\ref{ngs}. As a result of the fast temperature recovery and slow excess density propagation, a strong pressure perturbation is obtained along the magnetic field lines affected by the pellet source. The pressure gradient is both parallel and anti-parallel to the pellet trajectory (depending on which side of the pellet position is being considered). The localised pressure gradient to the right of the pellet position is parallel to the field line curvature and, therefore, may excite ballooning modes in the HFS.

\begin{figure}[!ht]
\centering
  \includegraphics[width=0.49\textwidth]{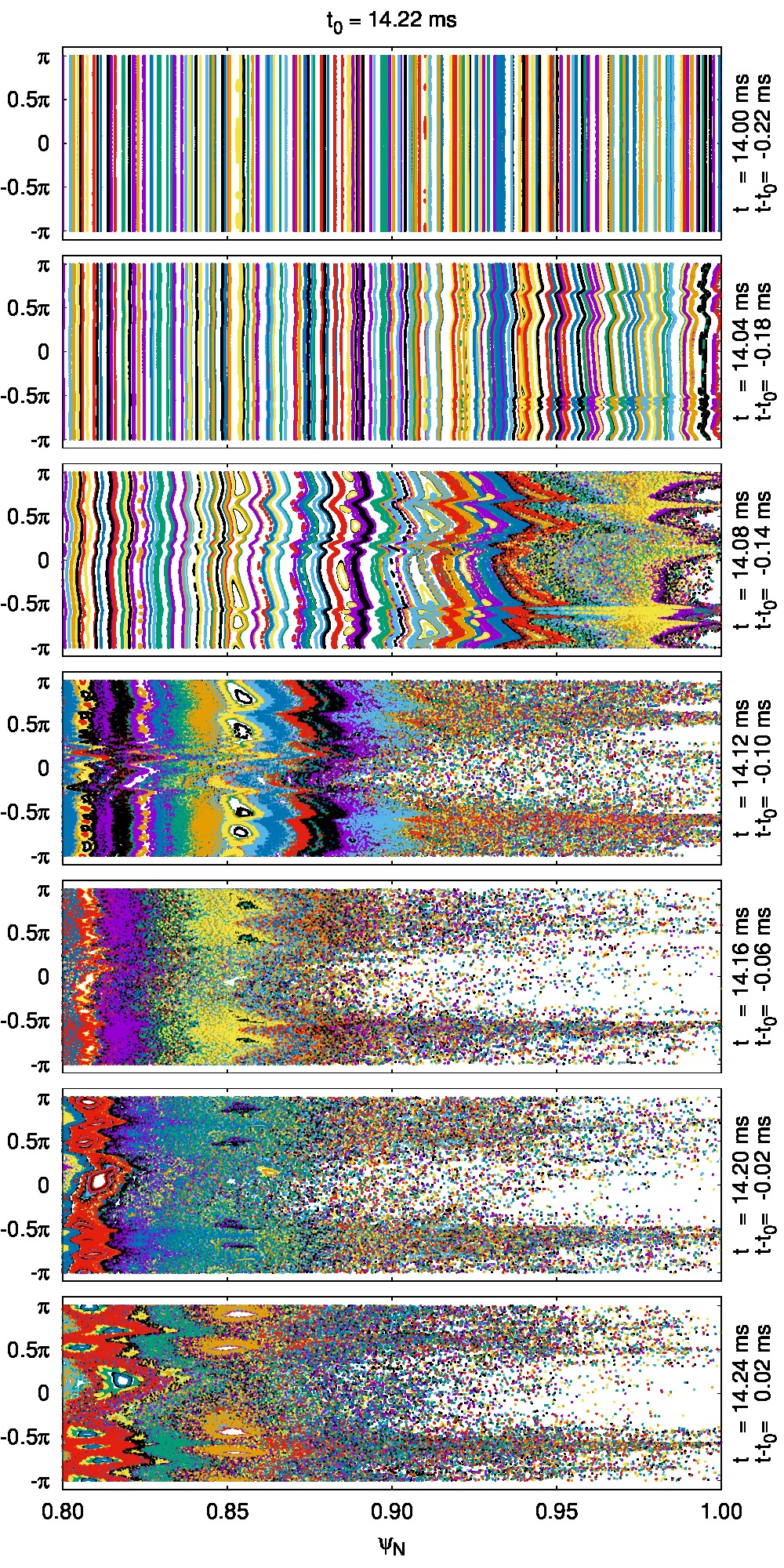}
  \caption{Seven Poincar\'e plots at different times with respect to the time of maximum $\Podiv$, ${t_0=14.22~\mathrm{ms}}$. From top to bottom, the chosen times are separated by ${0.04~\mathrm{ms}}$. The initially axisymmetric topology (top panel) becomes perturbed when the pellet crosses the separatrix (second panel), and later starts to reconnect (third panel). The ELM is then triggered (fourth panel), and ultimately the region with short connection length (white region) penetrates until ${\psin\approx0.87}$ (bottom panels).}
\label{fig:triggeredELM-poincare}
\end{figure}

The time of maximum outer divertor incident power is ${t_0=14.22~\mathrm{ms}}$. Seven different times with equal spacing $0.04~\mathrm{ms}$ are chosen with respect to the time of maximum outer divertor incident power ${t-t_0=-0.22:0.04:0.02~\mathrm{ms}}$ in order to analyse how the edge magnetic topology changes with pellet injection. For this pellet-triggered ELM, Poincar\'e plots at the times mentioned before are shown in fig.~\ref{fig:triggeredELM-poincare}. At ${t-t_0=-0.22~\mathrm{ms}}$, i.e. at the time of pellet injection, the edge magnetic field is axisymmetric. Forty microseconds later, as the pellet crosses the separatrix (see fig.~\ref{fig:triggeredELM-ped}), the magnetic topology outwards of ${\psin\approx0.90}$ has become non-axisymmetric, but no reconnection has yet taken place.

In the next time slice (${t-t_0=-0.14~\mathrm{ms}}$, i.e. ${t=t_\mathrm{inj.}+0.08~\mathrm{ms}}$) the pellet has reached the radial position of $\psin\approx0.96$. At this point, some magnetic reconnection has taken place and a stochastic edge is formed from the last closed flux surface until the pellet position. This time slice is the last to clearly represent the pellet-induced magnetic perturbation. The following time slices (${t\geq t_{inj.}+0.12}$ ms) feature a stochastic magnetic topology mostly caused by the MHD response to the pellet perturbation. This can be distinguished because the stochastic layer has penetrated significantly further inwards than the pellet has. Namely, at ${t=t_\mathrm{inj.}+0.12~\mathrm{ms}}$ the pellet location is ${\psin\approx0.95}$ while the stochastic region has penetrated until ${\psin\approx0.90}$. Similarly to the spontaneous ELM crash (fig.~\ref{fig:spontaneousELM-poincare}), field lines at $\Psi_N\gtrsim0.87$ exhibit a very short connection length to the divertor targets close to the time of maximum divertor heat flux.

\begin{figure}
\centering
  \includegraphics[width=0.49\textwidth]{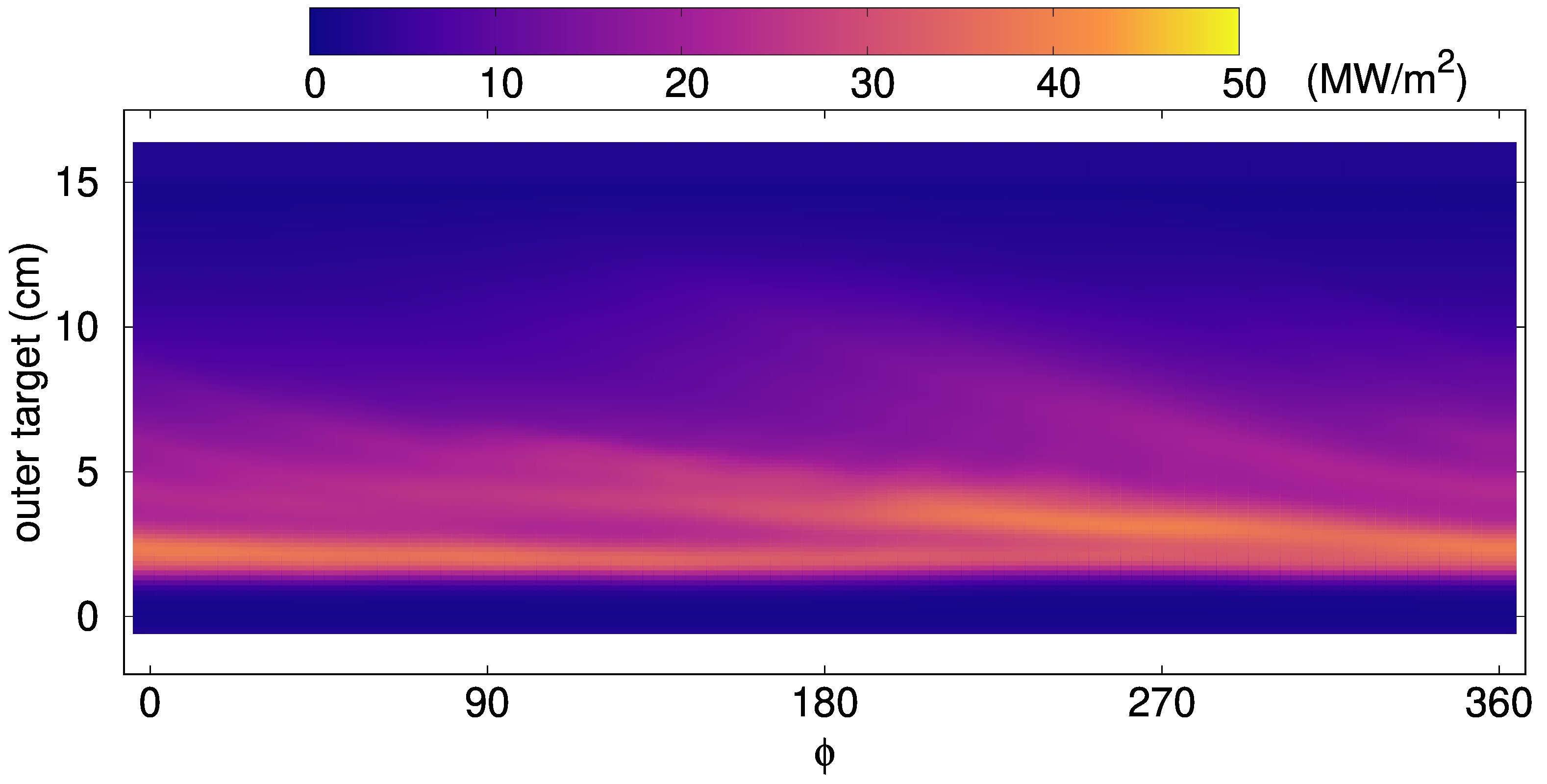} 
\caption{Outer divertor heat flux profile at the time of maximum $\Podiv$, $t_0=14.22~\mathrm{ms}$. The footprint corresponding to two main strike lines can be observed. The wetted area at this time point is ${A_\mathrm{wet}\approx0.65~\mathrm{m^2}}$, which is $\sim31~\%$ lower than the spontaneous ELM \texttt{Sp-318.1}.}
\label{fig:triggeredELM-heatflux}
\end{figure}

The energy expelled by the pellet-triggered ELM is then non-axisymmetrically deposited on the divertor targets. Similar to the spontaneous ELM shown before, the pellet-triggered ELM deposits more energy (roughly twice) onto the outer divertor than the inner divertor. This is in fact a feature of all the pellet-triggered ELMs described in the present work. Using the time of maximum outer divertor incident power, ${t_0=14.22~\mathrm{ms}}$, the heat flux onto the outer target at different toroidal angles is shown in fig.~\ref{fig:triggeredELM-heatflux}. This profile shows two distinct regions with large heat deposition: one located in the original strike-line position (${\sim2~\mathrm{cm}}$) and a secondary deposition region that varies in intensity depending on the toroidal angle (peaking in intensity at ${\phi\approx240^\circ}$). This non-axisymmetric secondary deposition region rotates slowly along the toroidal direction and is closely linked to lobe structures.

Comparing the instantaneous heat flux profiles along the toroidal angle for the spontaneous (fig.~\ref{fig:spontaneousELM-heatflux}) and pellet-triggered ELMs (fig.~\ref{fig:triggeredELM-heatflux}), it is clear that the wetted area for the latter is significantly reduced. As stated before, the wetted area at the time of maximum outer divertor incident power is $A_\mathrm{wet}\approx0.95~\mathrm{m^2}$ for the spontaneous ELM (\texttt{Sp-318.1}). On the other hand, for the pellet-triggered ELM analysed in this section (\texttt{Tr-08-14ms}), the wetted area at $t_0$ is reduced by approximately $31\%$ with respect to \texttt{Sp-318.1}, i.e.  $A_\mathrm{wet}\approx0.65~\mathrm{m^2}$. 

\subsection{Comparison of  \normalfont{\texttt{Sp-318}} and \normalfont{\texttt{Tr-08-14ms}}}\label{sidebyside}

\begin{figure}[!h]
\centering
  \includegraphics[width=0.49\textwidth]{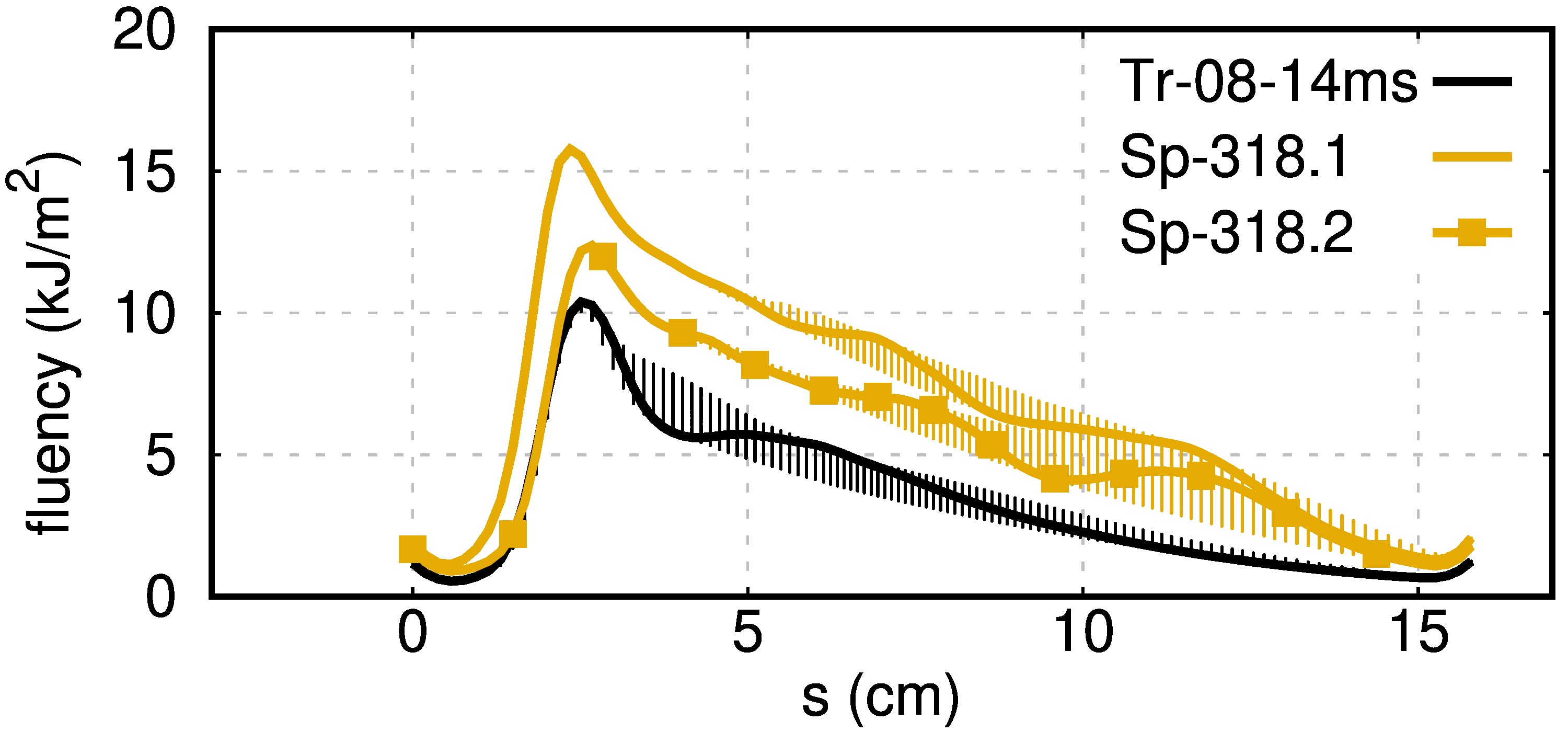}
  \caption{Outer divertor ELM fluency for two spontaneous ELMs (\texttt{Sp-318.1} and \texttt{Sp-318.2}) and an ELM triggered by a pellet containing $0.8\times10^{20}\mathrm{D}$ atoms (\texttt{Tr-08-14ms}). The wider wetted area associated to the spontaneous ELMs with respect to the triggered ELM can be observed.}
\label{fig:compare-fluency}
\end{figure}

The ELM outer target energy fluency, 
\begin{align}
    \varepsilon_\mathrm{target}(s,\phi) = \int_{t_\mathrm{ELM}} q(s,\phi,t) dt,  \label{eq:fluency}
\end{align}
for \texttt{Sp-318.1} (dark yellow lines), \texttt{Sp-318.2} (dark yellow lines with symbols), and \texttt{Tr-08-14ms} (black lines) is shown in fig.~\ref{fig:compare-fluency}. The full lines for each ELM correspond to the target fluency profile at ${\phi=0}$ and the small dots correspond to other toroidal angles. The figure clearly shows the reduction in wetted area between the pellet-triggered ELM and the spontaneous ELM that was described in the previous paragraph. 

The second spontaneous ELM, \texttt{Sp-318.2}, is included so that the fluency of a spontaneous ELM borne out of self-consistent seed perturbations can also be observed. There is a reduction in the peak target fluency for the triggered ELM with respect to both spontaneous ELMs: \texttt{Sp-318.1} and \texttt{Sp-318.2}. The second spontaneous ELM (full lines with symbols in fig.~\ref{fig:compare-fluency}) shows a lower peak fluency with respect to the first spontaneous ELM. This happens because \texttt{Sp-318.1} is borne out of noise-level perturbations, while \texttt{Sp-318.2} is borne out of self-consistent seed perturbations~\cite{Cathey2020}. The reduction of peak target fluency between \texttt{Tr-08-14ms} and \texttt{Sp-318.1} may be attributed to the lower ELM size of the triggered ELM (${\sim4.4\%}$) with respect to the spontaneous ELM (${\sim7.8\%}$). In the multi-machine scaling for the peak parallel energy fluency from Ref.~\cite{Eich2017A}, the ELM size goes into the scaling as $\Delta E_\mathrm{ELM}^{0.52}$. Scaling the peak target fluency of the spontaneous ELM (${\sim15.5~\mathrm{\frac{kJ}{m^2}}}$) with the pellet-triggered ELM size results in a peak value of ${15.5~\mathrm{\frac{kJ}{m^2}}\left(\frac{4.4}{7.8}\right)^{0.52}\approx11.5~\mathrm{\frac{kJ}{m^2}}}$, which is comparable to the peak fluency of ${\sim10.5~\mathrm{\frac{kJ}{m^2}}}$ measured for the pellet-triggered ELM. The remaining difference is attributed to the fact that the pedestal pressure right before the spontaneous ELM is larger than before the pellet-triggered ELM, as may be observed from fig.~\ref{fig:pressure}.

\begin{figure*}[!h]
\centering
  \includegraphics[width=0.49\textwidth]{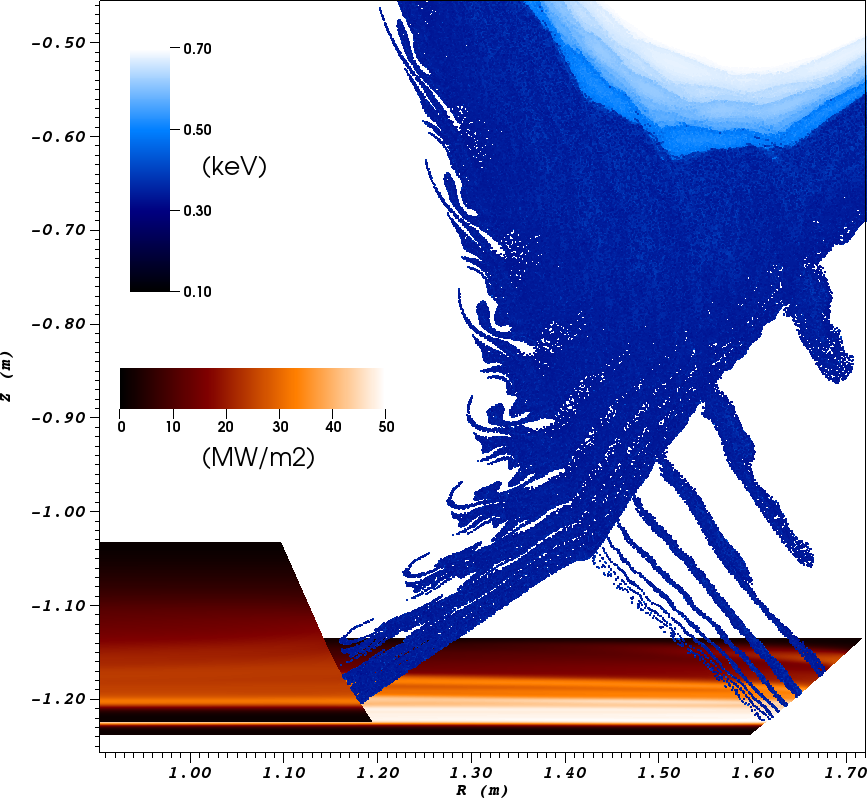}
  \includegraphics[width=0.49\textwidth]{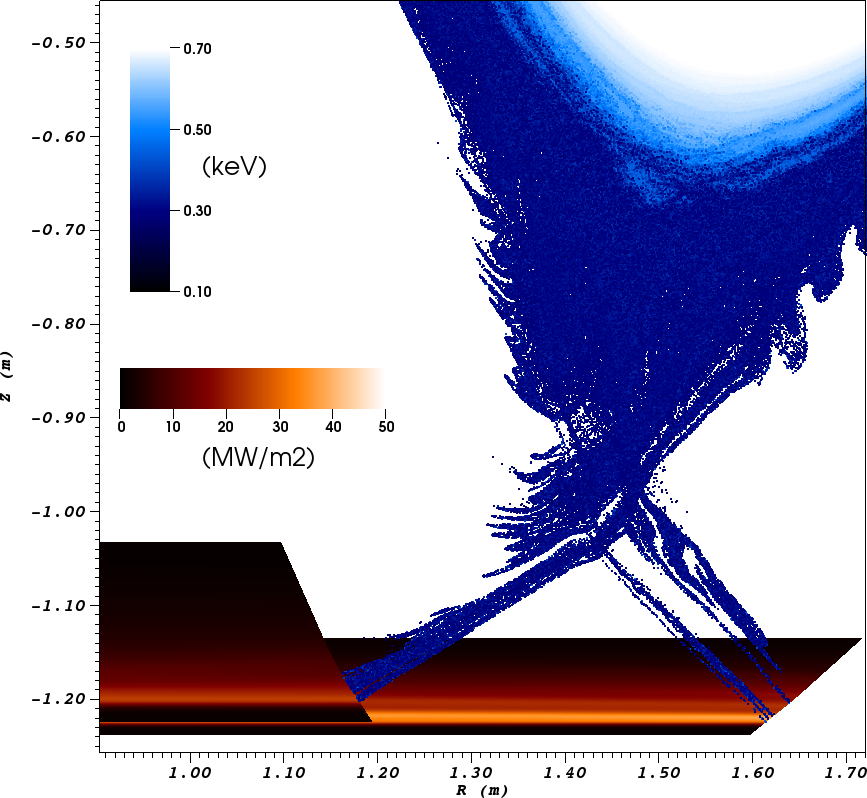} 
\caption{Heat flux to the inner and outer divertor targets ($\mathrm{MW/m^2}$) and real space Poincar\'e  plots with a colour scheme that reflects the temperature (keV) at the starting location of the associated magnetic field lines for (left) \texttt{Sp-318.1} and (right) \texttt{Tr-08-14ms} at the respective times of maximum ${\Podiv}$. The low number of secondary strike lines of the pellet-triggered ELM can explain the narrower wetted area when compared to the spontaneous ELM that splits the strike line into several secondary strike lines.}
\label{fig:compare-poincare}
\end{figure*}

A side-by-side comparison between \texttt{Sp-318.1} and \texttt{Tr-08-14ms} in terms of the heat fluxes to the inner and outer divertors, and of the non-axisymmetric topologies (portrayed by Poincar\'e maps in real space with a colour scale that reflects the temperature at the starting position of the magnetic field line associated with each point of the Poincar\'e map) is shown in fig.~\ref{fig:compare-poincare} at the times of maximum incident power to the outer divertor. The wider wetted area and higher peak values of the spontaneous ELM (left), with respect to the pellet-triggered ELM (right), can be clearly observed. It appears that the reason behind the wider wetted area in the spontaneous ELM is a larger number of secondary strike lines, which carry thermal energy from the bulk plasma, across the magnetic separatrix, towards the divertor targets. A slight difference is also observed in terms of the initial temperature of the magnetic field lines traced for the Poincar\'e map -- the stochastic region seems to reach further inwards for the spontaneous ELM than for the pellet-triggered ELM. 

\subsection{Mode structures}\label{modestructures}

\begin{figure*}[!h]
\centering
  \includegraphics[width=0.3\textwidth]{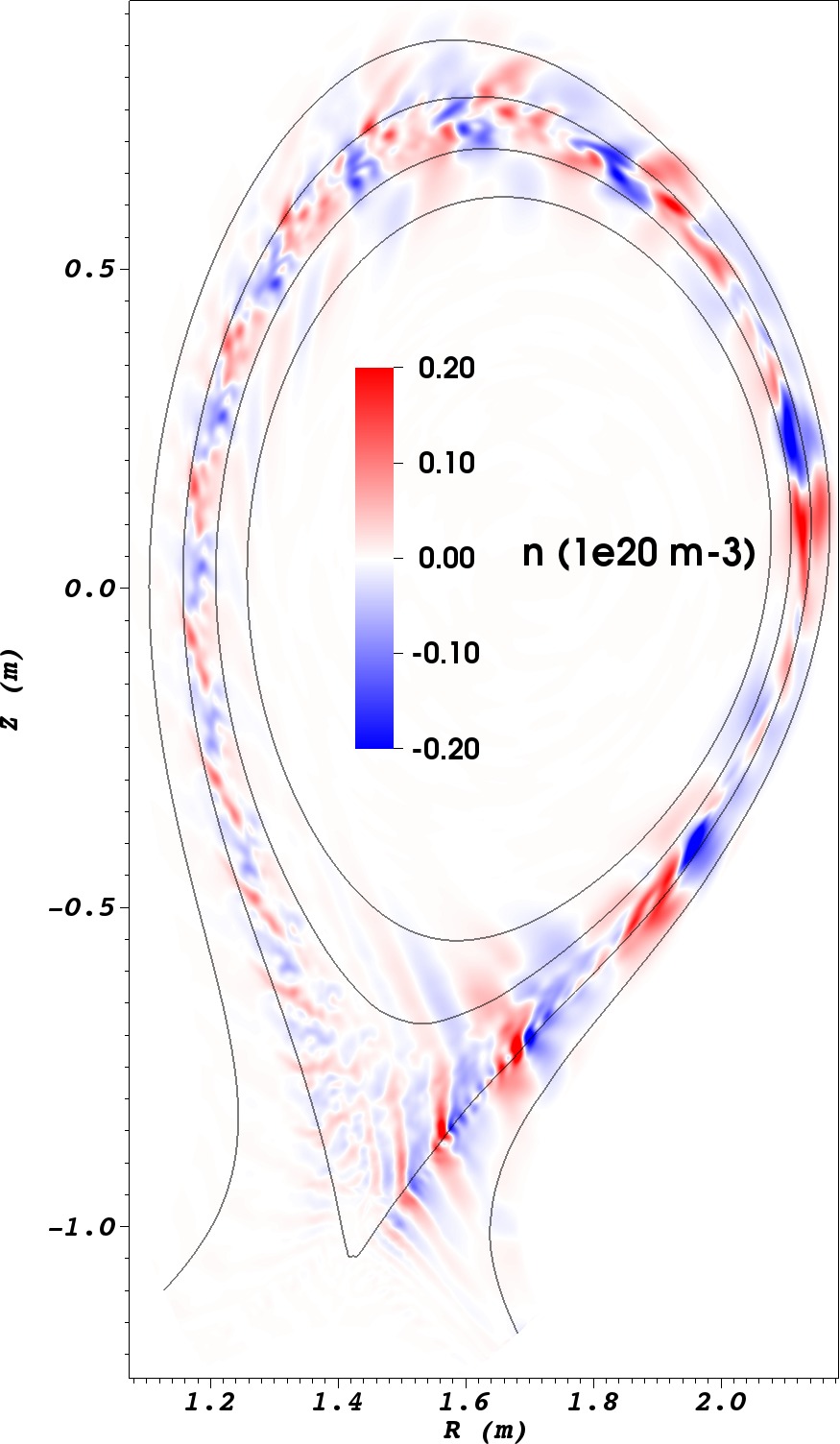}
  \includegraphics[width=0.3\textwidth]{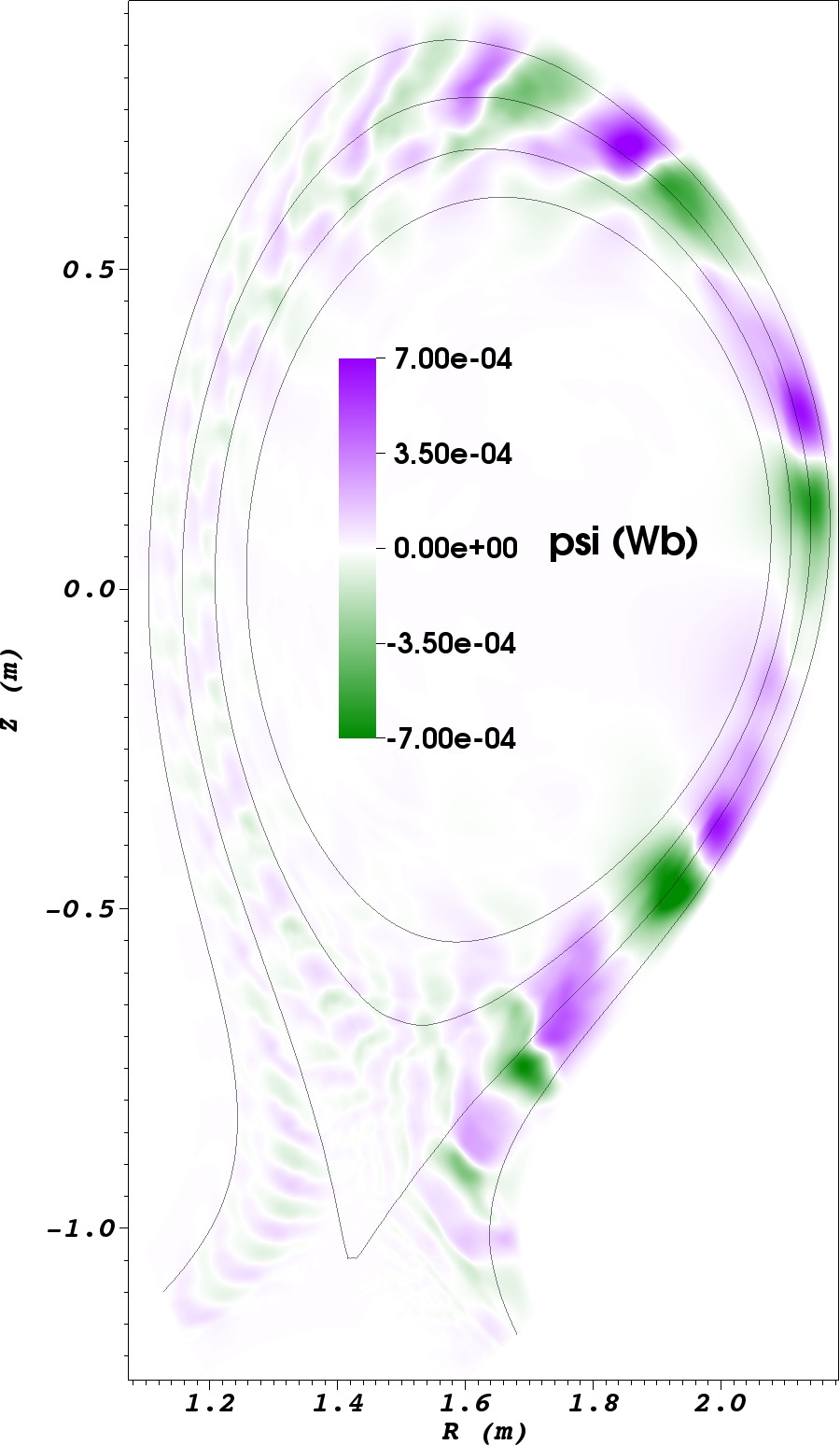}
  \includegraphics[width=0.3\textwidth]{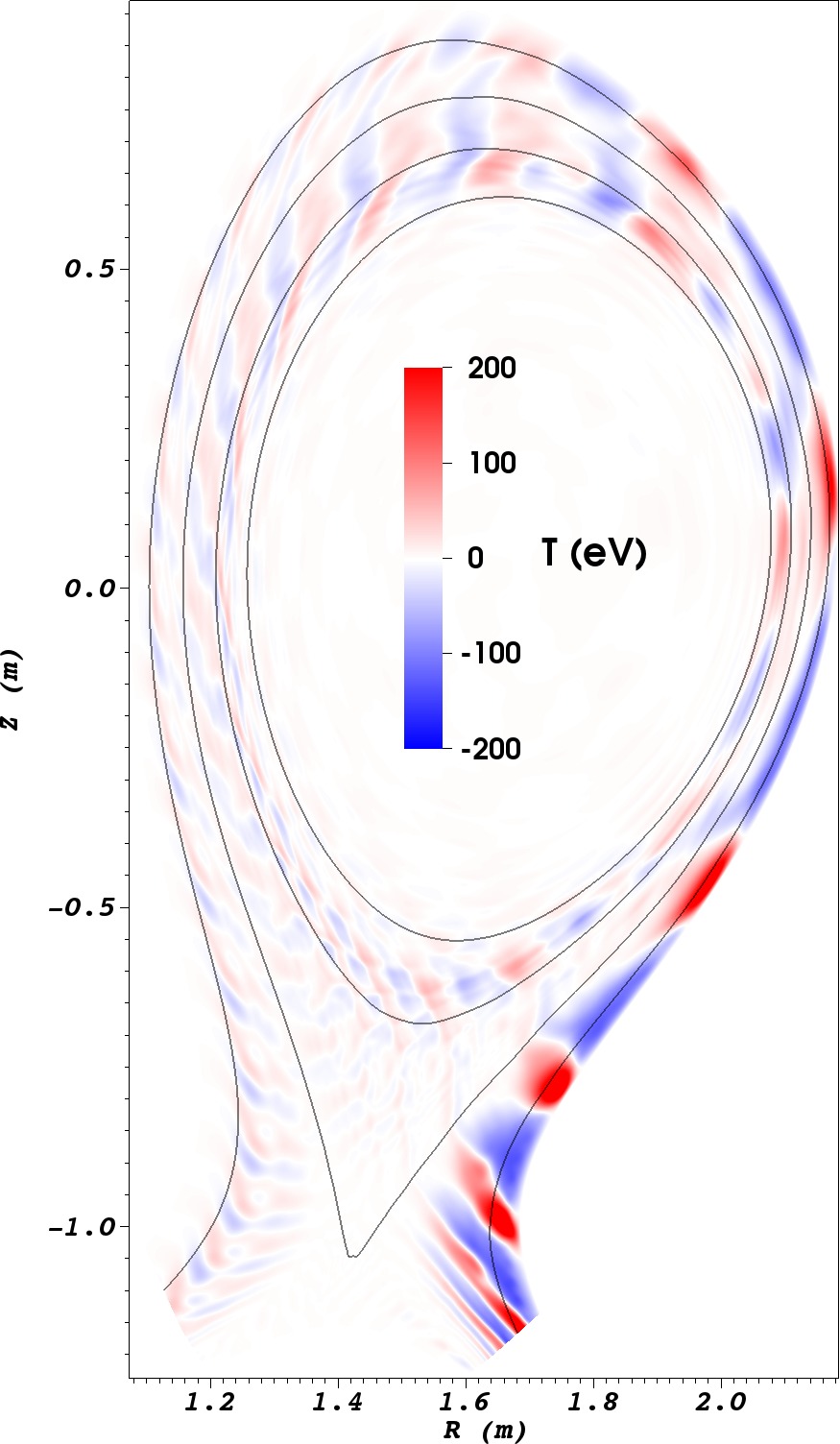}
  \includegraphics[width=0.3\textwidth]{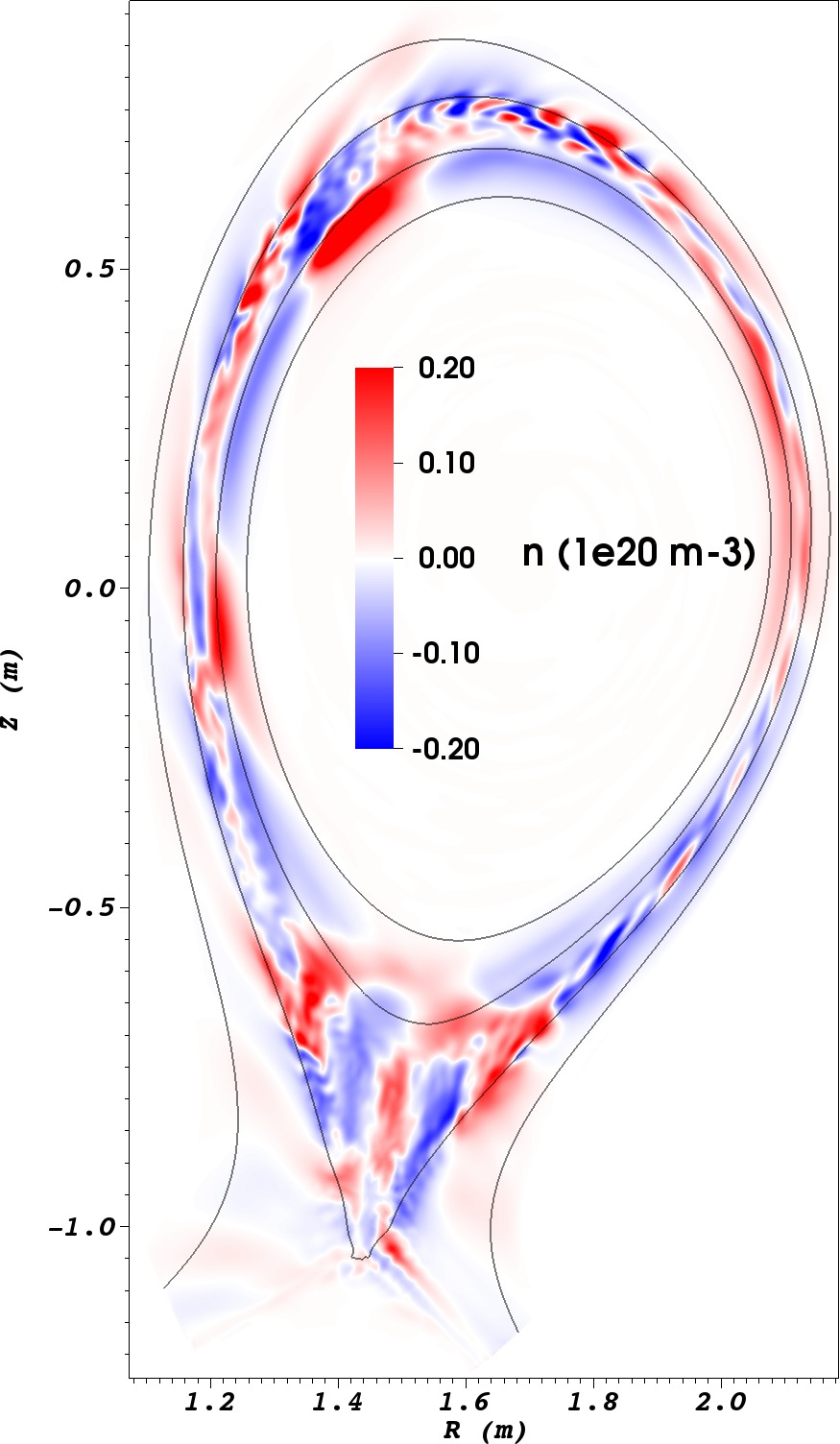}
  \includegraphics[width=0.3\textwidth]{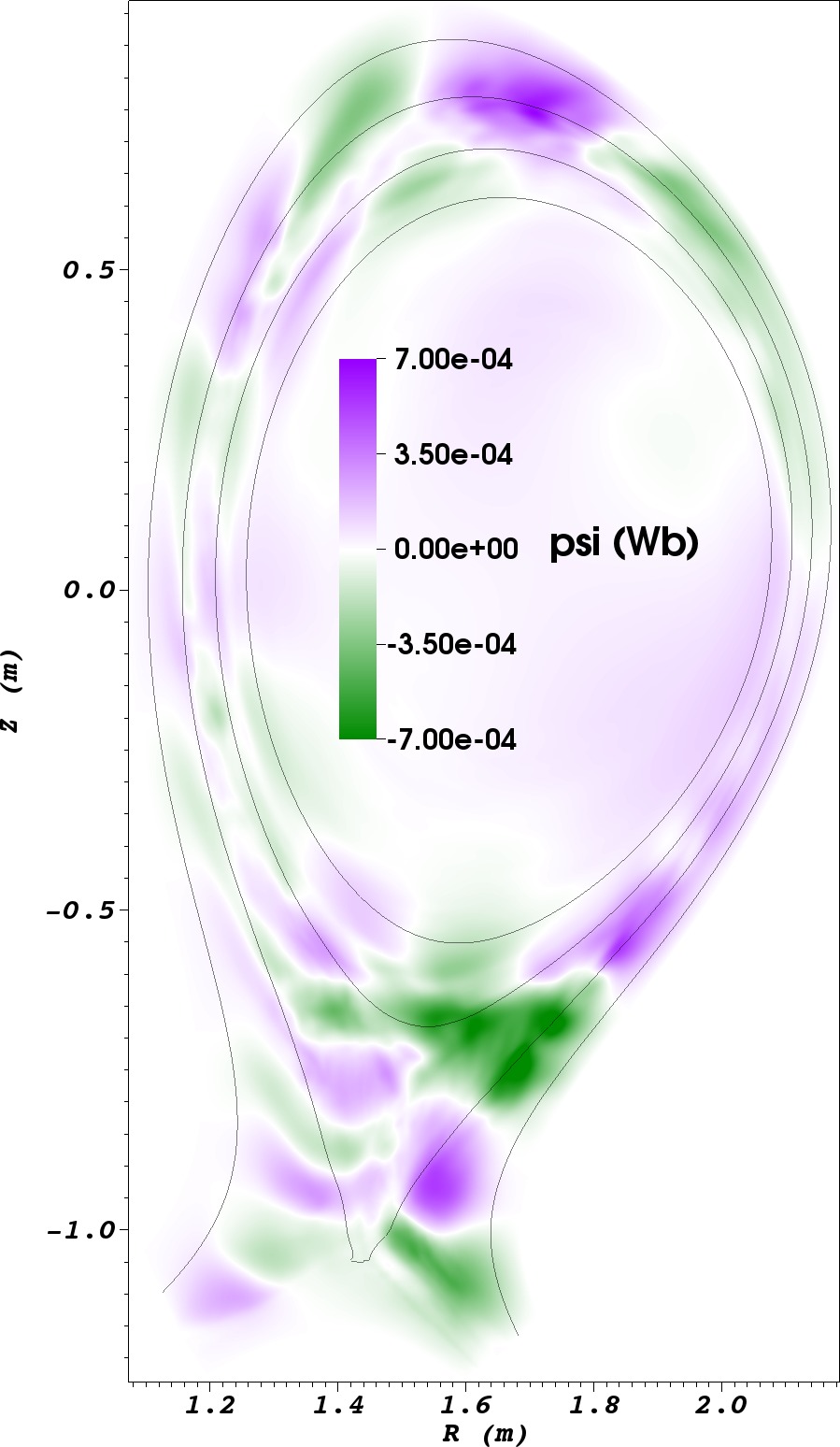}
  \includegraphics[width=0.3\textwidth]{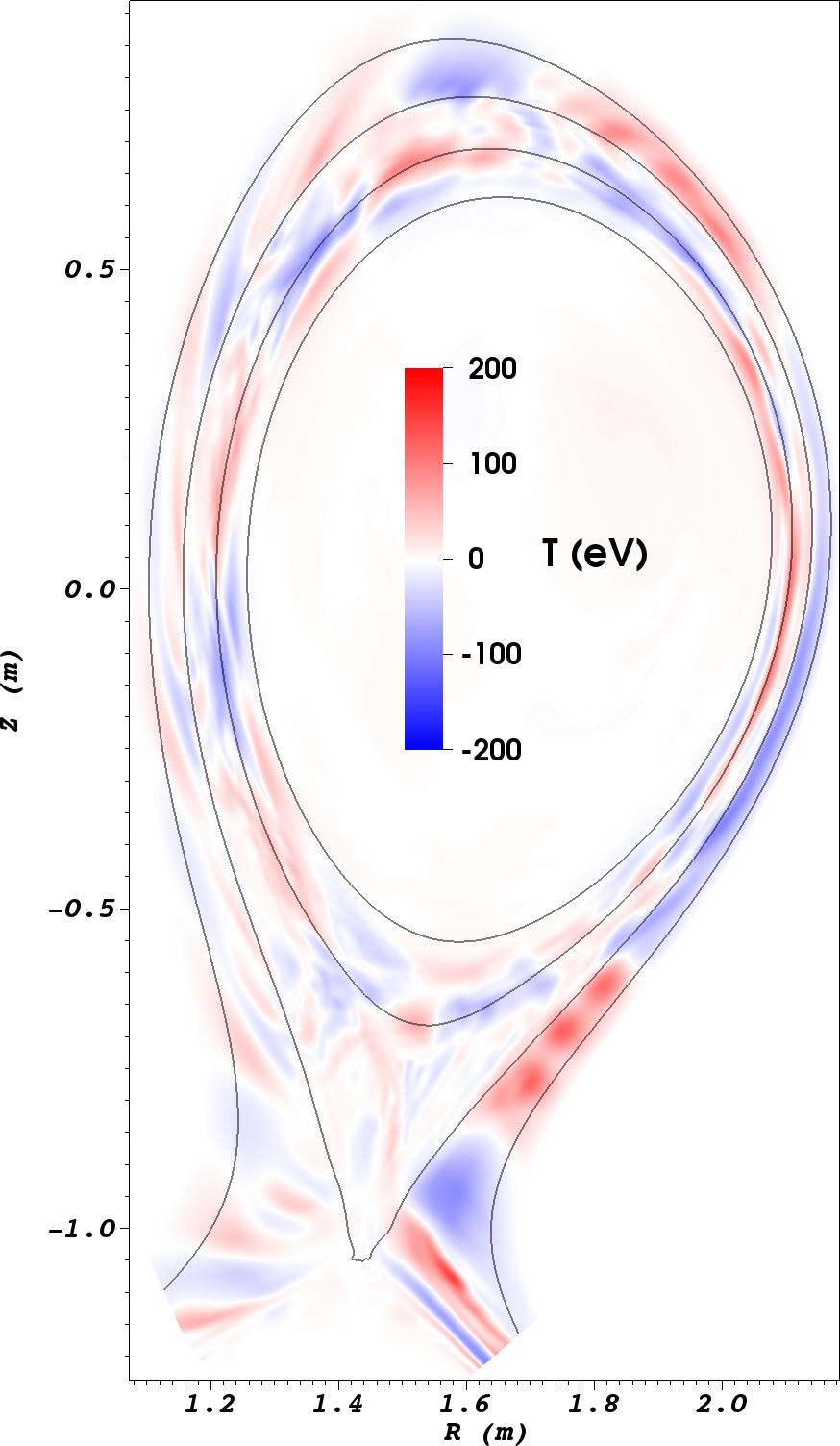}
  \caption{Density (in ${10^{20}~\mathrm{m^{-3}}}$), flux (in ${\mathrm{Wb}}$), and temperature (in ${\mathrm{eV}}$) perturbations for \texttt{Sp-318.1} (top row), and for \texttt{Tr-08-14ms} (bottom row), at the respective times of maximum ${\Podiv}$. Gray lines in all plots indicate surfaces with normalised poloidal flux of ${\psin=[0.8,0.9,1.0,1.1]}$. The mode structures for the spontaneous ELM feature peeling-ballooning modes more clearly than for the triggered ELM. The spatial dimensions of the modes are markedly larger for \texttt{Tr-08-14ms} than for \texttt{Sp-318.1}.}
  \label{fig:compare-perturbations}
\end{figure*}

In this subsection we present different quantities related to the structure of the modes involved in the ELM crashes described before, \texttt{Sp-318.1} and \texttt{Tr-08-14ms}. In particular, we pick the time point related to the peak incident power onto the outer divertor for each event (the same times as figs.~\ref{fig:spontaneousELM-heatflux} and~\ref{fig:triggeredELM-heatflux} for the spontaneous and pellet-triggered ELMs, respectively). The density (in ${10^{20}~\mathrm{m^{-3}}}$), poloidal magnetic flux (in ${\mathrm{Wb}}$), and the total plasma temperature (in ${\mathrm{eV}}$) for the spontaneous ELM and the pellet-triggered ELM are shown in the top and bottom rows of fig.~\ref{fig:compare-perturbations}. All plots also indicate, with gray lines, surfaces with different values of normalised poloidal flux ($\psin=[0.8,0.9,1.0,1.1]$). The perturbations in question relate to the non-axisymmetric component of the total quantities: temperature may be decomposed into an axisymmetric component ${T_{\mathrm{n}=0}}$ and a non-axisymmetric component ${T_{\mathrm{n}>0}}$ to form the total temperature ${T=T_{\mathrm{n}=0}+T_{\mathrm{n}>0}}$, where ${\mathrm{n}}$ is the toroidal mode number.

The spontaneous ELM \texttt{Sp-318.1} (top row), clearly features peeling-ballooning structures inside the last closed flux surface. Ballooning structures predominantly on the LFS are observed in all quantities, but particularly in the perturbations to the poloidal magnetic flux (centre plot). The spontaneous ELM clearly features large temperature fluctuations on the scrape-off layer (SOL), which maintain the structure observed in the Poincar\'e plot of fig.~\ref{fig:compare-poincare} (left). Interestingly, the pellet-triggered ELM shows faint hints of peeling-ballooning structures (the most obvious indications are shown in the HFS density perturbations enclosed between ${\psin\approx0.9}$ and ${1.0}$). It is unclear whether this is due to the large pellet-induced perturbation, which may alter or mask the natural structure of said modes, or due to fundamentally different perturbations related to the pellet-triggered ELM. 

Several differences may be observed in the spatial structures of the modes related to the spontaneous and pellet-triggered ELMs. One of the most obvious differences is the large density perturbation created by the pellet ablation in fig.~\ref{fig:compare-perturbations} (bottom, left). Overall, the density perturbations present in the volume enclosed between ${\psin\approx0.8}$ and ${1.0}$ are larger for \texttt{Tr-08-14ms} than for \texttt{Sp-318.1}. The large (in size and amplitude) density perturbation induced by the pellet ablation is observed to influence the size of the perturbations related to the ELM crash, i.e. the density perturbations observed for the pellet-triggered ELM appear to be larger than those for the spontaneous ELM. This is perhaps most clearly evidenced by comparing the plots of the magnetic flux perturbations (top and bottom centre plots). The temperature perturbations obtained for the pellet-triggered ELM appear to be generally weaker than those for the spontaneous ELM. These perturbations also demonstrate the structures observed in fig.~\ref{fig:compare-poincare} (right). Finally, it is interesting to note that while the spontaneous ELM features weak (density, flux, and temperature) perturbations in the confined region near the magnetic X-point, the pellet-triggered ELM shows large perturbations (particularly of density and flux) in this region.

\section{Comparison between spontaneous and pellet-triggered ELMs} \label{compare}

Having shown in detail the differing dynamics of a representative spontaneous and pellet-triggered ELM, the attention is now turned towards comparing several key quantities between more simulations of spontaneous and pellet-triggered ELMs. At first, in subsection~\ref{wth-losses}, a comparison in terms of the thermal energy losses caused by the different types of ELMs is described. Later, an analysis of the toroidal mode spectrum is discussed in subsection~\ref{spectrum}. Finally, in subsection~\ref{heatflux}, we study and detail how the respective heat fluxes and resulting energy fluencies compare.

\subsection{ELM-induced thermal energy losses}\label{wth-losses}

\begin{figure}
  \centering
  \includegraphics[width=0.49\textwidth]{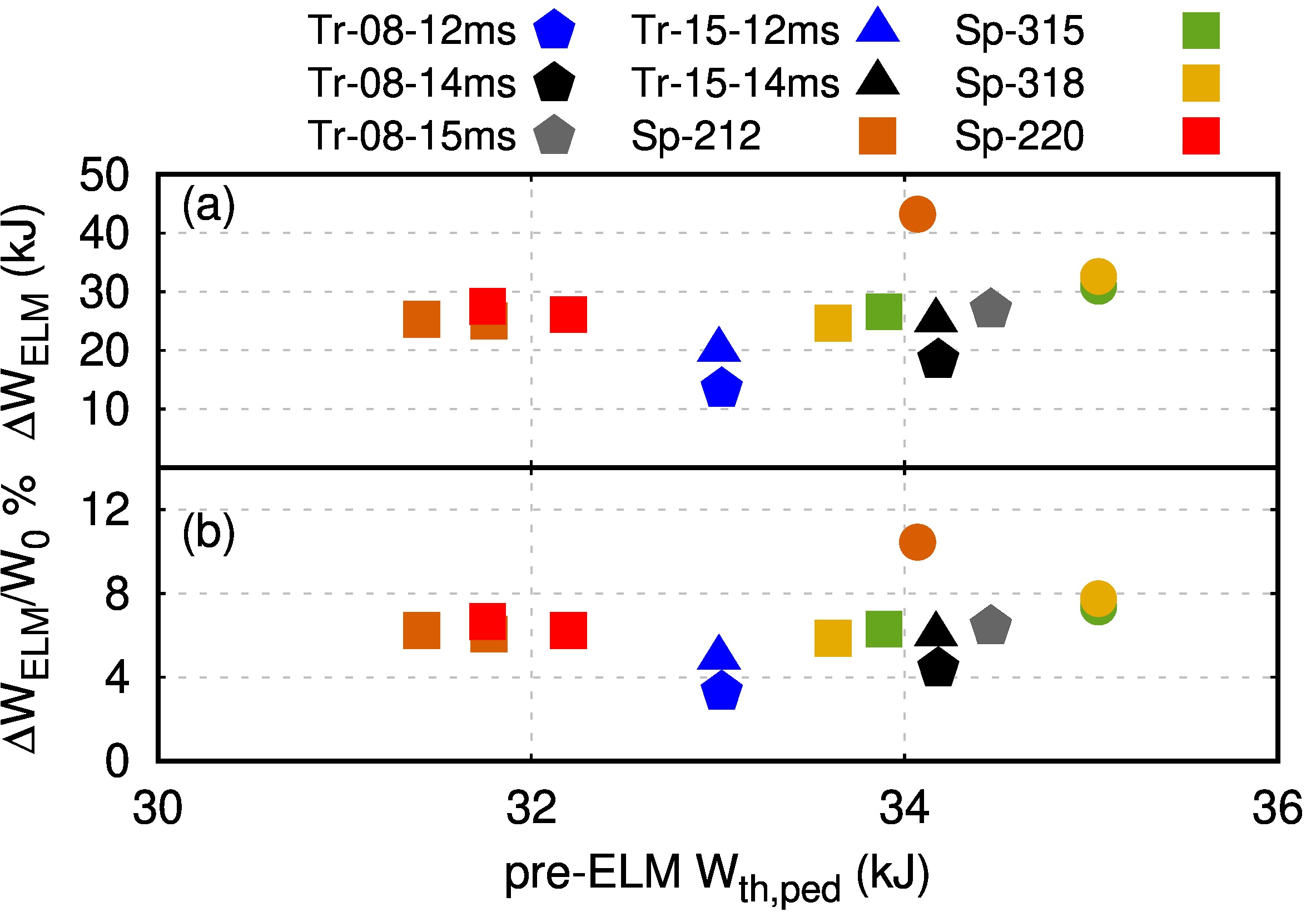}
  \caption{Energy losses (a) and relative ELM size (b) of spontaneous (circles denote the first simulated ELM in a series, while squares represent the subsequent ELMs) and pellet-triggered ELMs (pentagons for pellets containing ${0.8\times10^{20}~\mathrm{D}}$ atoms and triangles for pellets containing ${1.5\times10^{20}~\mathrm{D}}$ atoms). Spontaneous ELMs borne out of self-consistent seed perturbations have similar absolute and relative ELM sizes. Early enough injection can reduce the ELM size with respect to spontaneous ELMs.}
  \label{fig:compare-energy-losses}
\end{figure}

The energy released by the different ELMs (${\Delta W_\mathrm{ELM}}$) and the relative ELM size (${\Delta E_\mathrm{ELM}}$) are plotted against the pre-ELM pedestal stored energy (pre-ELM ${\mathrm{W_{th,ped}}}$ is obtained with a volume integral from $\psin=0.9$ to the separatrix) in fig.~\ref{fig:compare-energy-losses} (a) and (b), respectively. In said figure, the pentagons represent ELMs triggered by pellets containing $0.8\times10^{20}\mathrm{D}$ atoms, while the triangles correspond to pellets with $1.5\times10^{20}\mathrm{D}$ atoms. Squares and circles represent spontaneous ELMs, and different colours are used for simulations with different sets of toroidal mode numbers included, e.g. dark yellow for ${\mathrm{n}=0-3-18}$. The circles denote the first ELM simulated for a given series of ELM crashes, while the squares denote the subsequent ELMs, which are borne out of self-consistent seed perturbations (the reason for this differentiation is described in section~\ref{mode-numbers}).

As a first observation, we point out the fact that all spontaneous ELMs borne out of self-consistent seed perturbations, i.e., represented by squares in fig.~\ref{fig:compare-energy-losses}, have similar absolute and relative ELM sizes. The first spontaneous ELMs in a given series of simulated ELMs (circles in fig.~\ref{fig:compare-energy-losses}) have larger thermal energy losses, with respect to the other spontaneous ELMs, because they are borne out of lower amplitude seed perturbations~\cite{Cathey2020}. 

Considering the pellet-triggered ELMs alone, it is observed that pellets (small and large) injected at an earlier time (lower pre-ELM ${\mathrm{W_{th,ped}}}$) result in smaller ELM losses with respect to similar-sized pellets injected at a later time (larger pre-ELM ${\mathrm{W_{th,ped}}}$). For example, the blue pentagons have a smaller absolute, and relative, ELM size than their black and gray counterparts. For a given injection timing, the larger pellets induce a slightly larger triggered ELM size (triangles lie above pentagons for constant pre-ELM ${\mathrm{W_{th,ped}}}$). 

Comparing the spontaneous and pellet-triggered ELMs reveals only a small difference in ELM size, and such difference is most pronounced when comparing the earliest pellet injection. With the large pellet, ELMs may be triggered at even earlier injection times (${t_\mathrm{inj}\geq8~\mathrm{ms}}$), but the small pellet is able to trigger ELMs only if ${t_\mathrm{inj}\geq12~\mathrm{ms}}$ (further details found in Ref.~\cite{Futatani2020}). Finally, it is worth pointing out that the small pellet injected at the latest time (only ${1~\mathrm{ms}}$ before the spontaneous ELM crash would take place) triggers an ELM with a comparable size to the spontaneous ELMs simulated in these studies. The small differences between spontaneous and pellet-triggered ELMs appear to be in agreement with experimental observations at AUG~\cite{Lang_2014}.

We observe that spontaneous ELMs typically expel more thermal energy than pellet-triggered ELMs, except when the pellet is injected very briefly before the spontaneous ELM would appear, i.e. at roughly the same pedestal conditions. For the pellet-triggered ELMs simulated, we also observe that large pellets induce slightly larger thermal energy losses than ELMs triggered by small pellets injected at the same time.

\subsection{Toroidal mode spectrum}\label{spectrum}

\begin{figure}
\centering
  \includegraphics[width=0.49\textwidth]{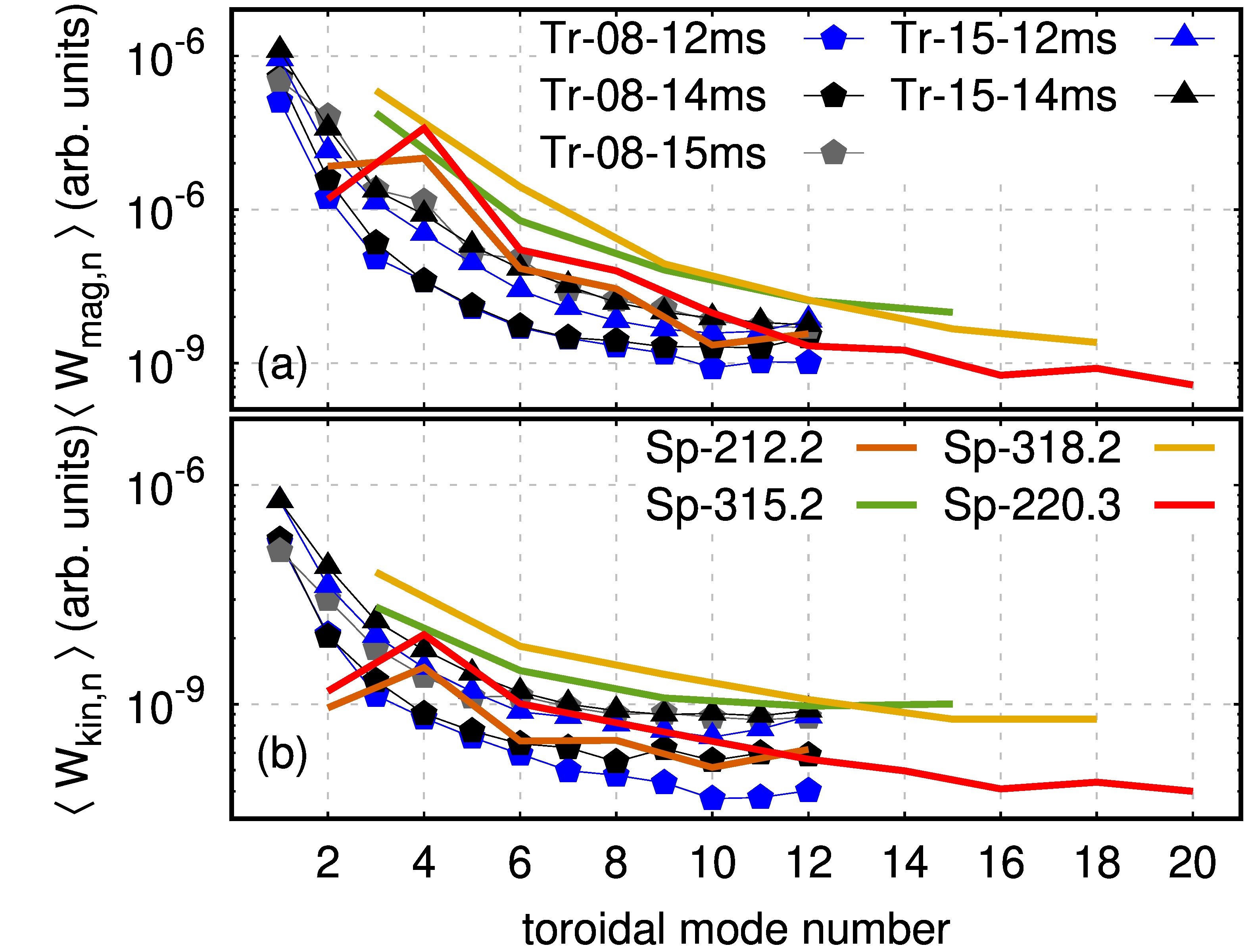} 
\caption{Time averaged magnetic (top) and kinetic (bottom) energy spectrum for spontaneous (full lines) and pellet-triggered ELMs (symbols). The average is performed over ${0.5~\mathrm{ms}}$ starting from ${t(W=W_{\mathrm{mag,max}})-0.1~\mathrm{ms}}$.}
\label{fig:spectra}
\end{figure}

Time-averaged spectra of the perturbations associated to the spontaneous and pellet-triggered ELMs are shown in fig.~\ref{fig:spectra} for the different toroidal mode numbers. The averaging is performed over ${0.5~\mathrm{ms}}$ starting ${0.1~\mathrm{ms}}$ before the time at which ${\sum_{\mathrm{n}>1}^{\mathrm{n_{max}}}W_{\mathrm{mag,n}}}$ is maximised. Excluding ${\mathrm{n}=1}$ from the summation is done in order to separate, to some extent, the pellet-induced perturbation from the pellet-triggered ELM crash.

For all the pellet-triggered ELMs, the most energetic perturbation is the ${\mathrm{n}=1}$, followed by the next low ${\mathrm{n}}$ modes. Such observation appears to be in agreement with experimental observations at JET~\cite{Poli_2010} and is consistent with previous pellet-triggered ELM simulations with JOREK~\cite{Futatani2014,Futatani2019}. The energy spectrum of the pellet-triggered ELMs is broad since the pellet-induced perturbation is described by low-n modes, and it excites high-n ballooning modes (as described in section~\ref{triggered}). On the other hand, for the spontaneous ELMs simulated for this study, the most unstable toroidal mode numbers during the ELM crash are ${\mathrm{2,3,4}}$ (consistent with experimental observations at AUG~\cite{Mink2017}), and the broad mode spectrum is explained by the three-wave coupling of low-to-high toroidal mode numbers~\cite{Krebs2013}. For spontaneous ELM simulations which include the ${\mathrm{n}=1}$ mode (not shown), the most energetic perturbation is not the ${\mathrm{n}=1}$.

\begin{figure}
\centering
  \includegraphics[width=0.49\textwidth]{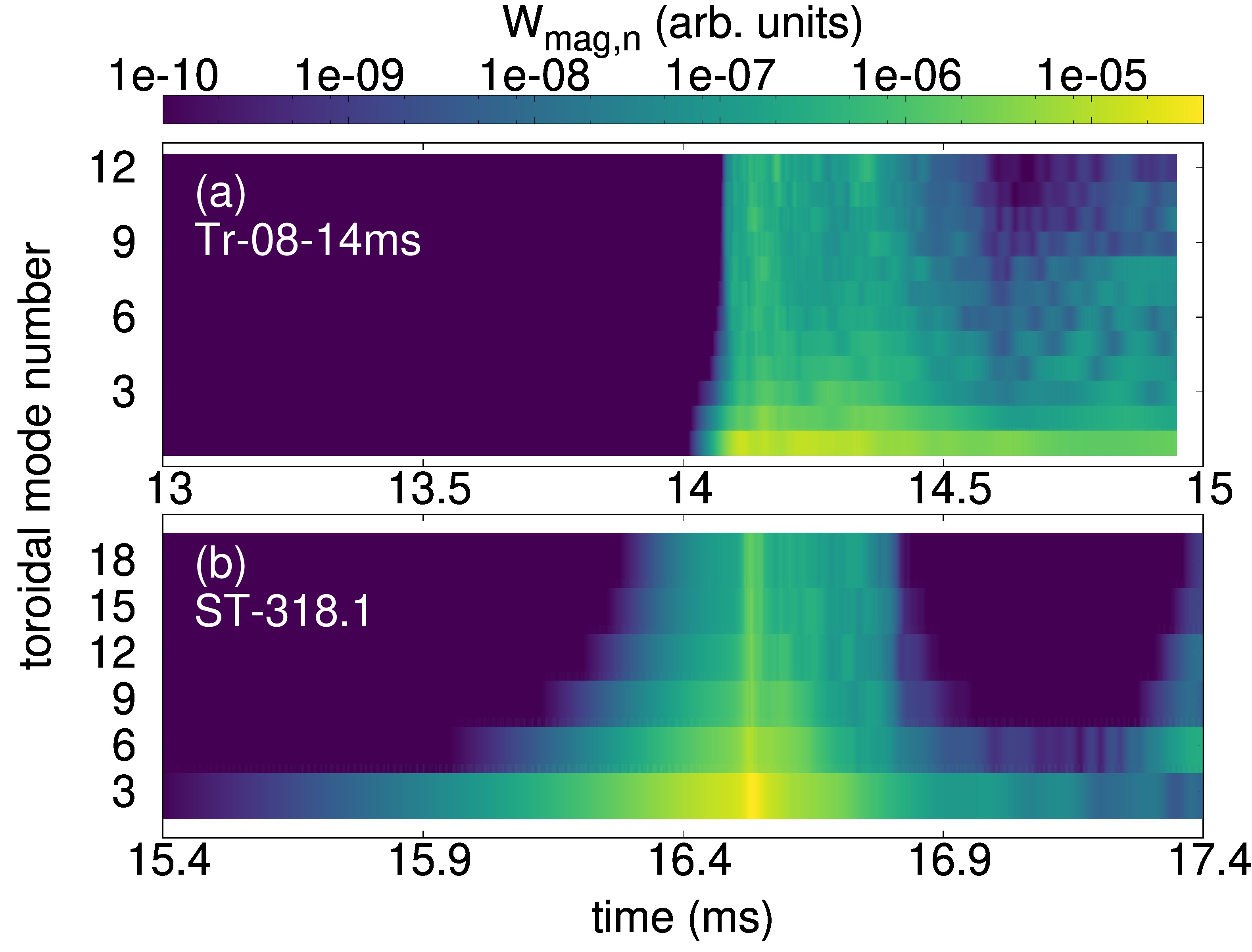} 
\caption{Time-evolving magnetic energy spectrum for \texttt{Tr-08-14ms} (a) and \texttt{Sp-318.1} (b). The time ranges chosen correspond to those of figs.~\ref{fig:spontaneousELM} and~\ref{fig:triggeredELM}, and the colour map is in logarithmic scale.}
\label{fig:compare-spectra}
\end{figure}

In order to show a direct comparison between pellet-triggered and spontaneous ELMs, a comparison between two time-evolving spectra is presented. For the individual ELMs, we choose the same ELMs as those for section~\ref{elms} (\texttt{Sp-318.1} and \texttt{Tr-08-14ms}). In fig.~\ref{fig:compare-spectra} we show the time-evolving magnetic energy spectrum, which highlights different dynamics for these two different types of events. In particular, the precursor phase of the spontaneous ELM (fig.~\ref{fig:compare-spectra}(b) is clearly visible for roughly ${1~\mathrm{ms}}$ prior to the ELM onset. On the other hand, the energy of the non-axisymmetric perturbations related to the pellet-induced perturbation and pellet-triggered ELM become abruptly excited roughly ${0.1~\mathrm{ms}}$ after the pellet is injected. These different temporal dynamics can also be clearly inferred from figs.~\ref{fig:spontaneousELM-poincare} and~\ref{fig:triggeredELM-poincare}. It is observed that the ELM duration is shorter for the spontaneous ELM than for the pellet-triggered ELM.

\subsection{Heat-flux and energy fluency comparison}\label{heatflux}

\begin{figure}
\centering
  {\includegraphics[width=0.49\textwidth]{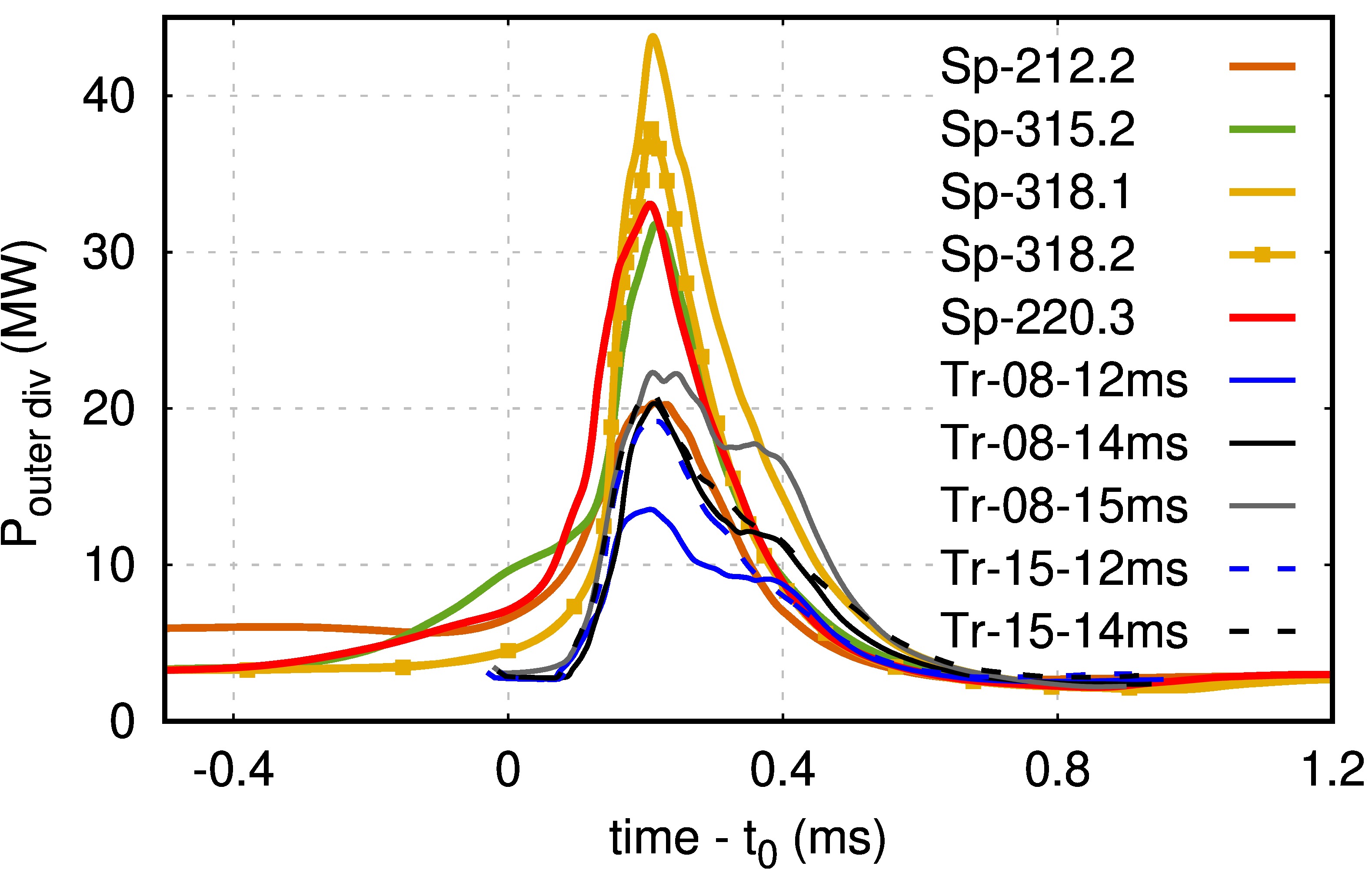} 
  \caption{Power incident on the outer divertor for several pellet-triggered ELMs (blue, black, and gray) and spontaneous ELMs. The temporal dynamics for simulations with ${\mathrm{n}=0-2-12}$ are too slow due to too low toroidal resolution. Simulations with ${\mathrm{n_{max}}\geq15}$ are properly converged.}
  \label{fig:pdiv}}
\end{figure}

An important metric for the comparison between the simulated spontaneous and pellet-triggered ELMs is the dynamical heat deposition profiles. As it was mentioned in section~\ref{spontaneous}, the splitting of the heat load between inner and outer divertor does not match experimental observations due to the simplified SOL modelling. Nevertheless, it is still of interest to analyse the differences and similarities between divertor heat fluxes for spontaneous and pellet-triggered ELMs. The outer divertor incident power,  ${\Podiv}$, is plotted for several spontaneous and triggered ELMs in fig.~\ref{fig:pdiv}. The time axis has been shifted such that the maximum of ${\Podiv}$ is at ${0.2~\mathrm{ms}}$ for all ELMs. The importance of keeping a sufficiently high toroidal resolution is highlighted by the peak divertor incident power shown between different spontaneous ELMs. In particular, \texttt{Sp-212.2} shows a peak divertor incident power roughly half that of \texttt{Sp-315.2} or \texttt{Sp-318.2}\footnote{We consider the simulations with ${\mathrm{n}=0-3-18}$ to be converged because they show only $\sim3\%$ lower peak outer divertor incident power with respect to simulations with ${\mathrm{n}=0-3-30}$.}. If we compare the pellet-triggered ELMs with \texttt{Sp-212.2} we observe very similar peak divertor incident powers. However, if we compare the pellet-triggered ELMs with the more realistic spontaneous ELMs (those with sufficient toroidal resolution), we observe a clear reduction in the peak divertor incident power by means of pellet injection. Indeed the latter conclusion is the one that should be drawn. 

At this stage it is worth pointing out that we have performed one simulation comparable to \texttt{Tr-08-14ms}, but with even higher toroidal mode numbers (${\mathrm{n}=0-1-18}$), and we observe a small (${\sim7\%}$) reduction of the peak divertor incident power with respect to the value in fig.~\ref{fig:pdiv}. Therefore, we consider the pellet-triggered ELM simulations presented in this paper to be properly converged. A more systematic convergence study, which would consider even higher toroidal resolution, is not affordable for us at present. 

From fig.~\ref{fig:pdiv} it can be observed that injecting a pellet with ${0.8\times10^{20}\mathrm{D}}$ atoms at ${12~\mathrm{ms}}$ leads to a much reduced peak divertor incident power with respect to the other four pellet-triggered ELMs. These remaining four pellet-triggered ELMs have similar peak ${\Podiv}$, which is weaker than all the spontaneous ELMs (excluding \texttt{Sp-212.2} which is not yet properly converged). This significant reduction in the peak divertor incident power does not translate directly to the target fluency (fig.~\ref{fig:compare-fluency}) because of the narrower deposition area intrinsic to the pellet-triggered ELMs.

\begin{figure}
\centering
  {\includegraphics[width=0.49\textwidth]{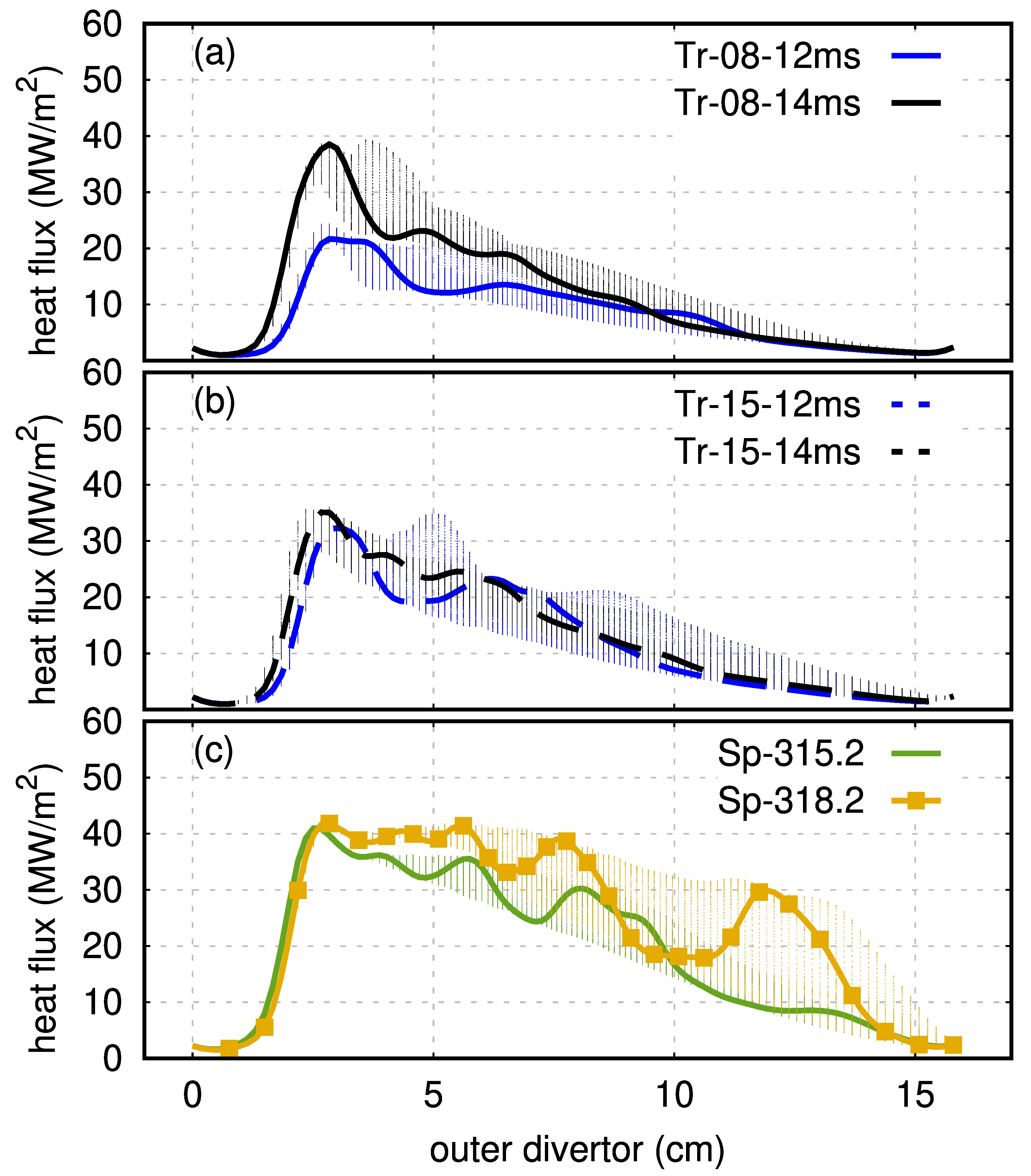} 
\caption{Outer divertors heat flux profile at ${t=t\left(max(P_\mathrm{out,div})\right)}$ for pellet-triggered ELMs (small and large pellets injected at $12$ and $14~\mathrm{ms}$) and spontaneous ELMs. Pellet-triggered ELMs (a) and (b) show a narrower deposition profile with respect to spontaneous ELMs (c).}
\label{fig:qatpmax}}
\end{figure}

We choose the time of maximum outer divertor incident power to clearly show the heat flux profile narrowing for the pellet-triggered ELMs. Figure~\ref{fig:qatpmax} shows the heat flux profile at ${\phi=0}$ (pellet injection angle) as solid or dashed lines, and at all other toroidal angles as small coloured points. This is done for six different ELMs: four pellet-triggered ELMs (triggered by a small pellet (a) and by a large pellet (b)) and two spontaneous ELMs (c). The pellet-triggered ELMs show much narrower heat flux deposition profiles with respect to the spontaneous ELMs. It is interesting to note that the non-axisymmetry observed for the pellet-triggered ELMs is most prominent in the vicinity of the maximum heat flux (${s\approx3~\mathrm{cm}}$), while for the spontaneous ELMs it is most evident away from the maximum. 

\begin{figure}
    \centering
    \includegraphics[width=0.49\textwidth]{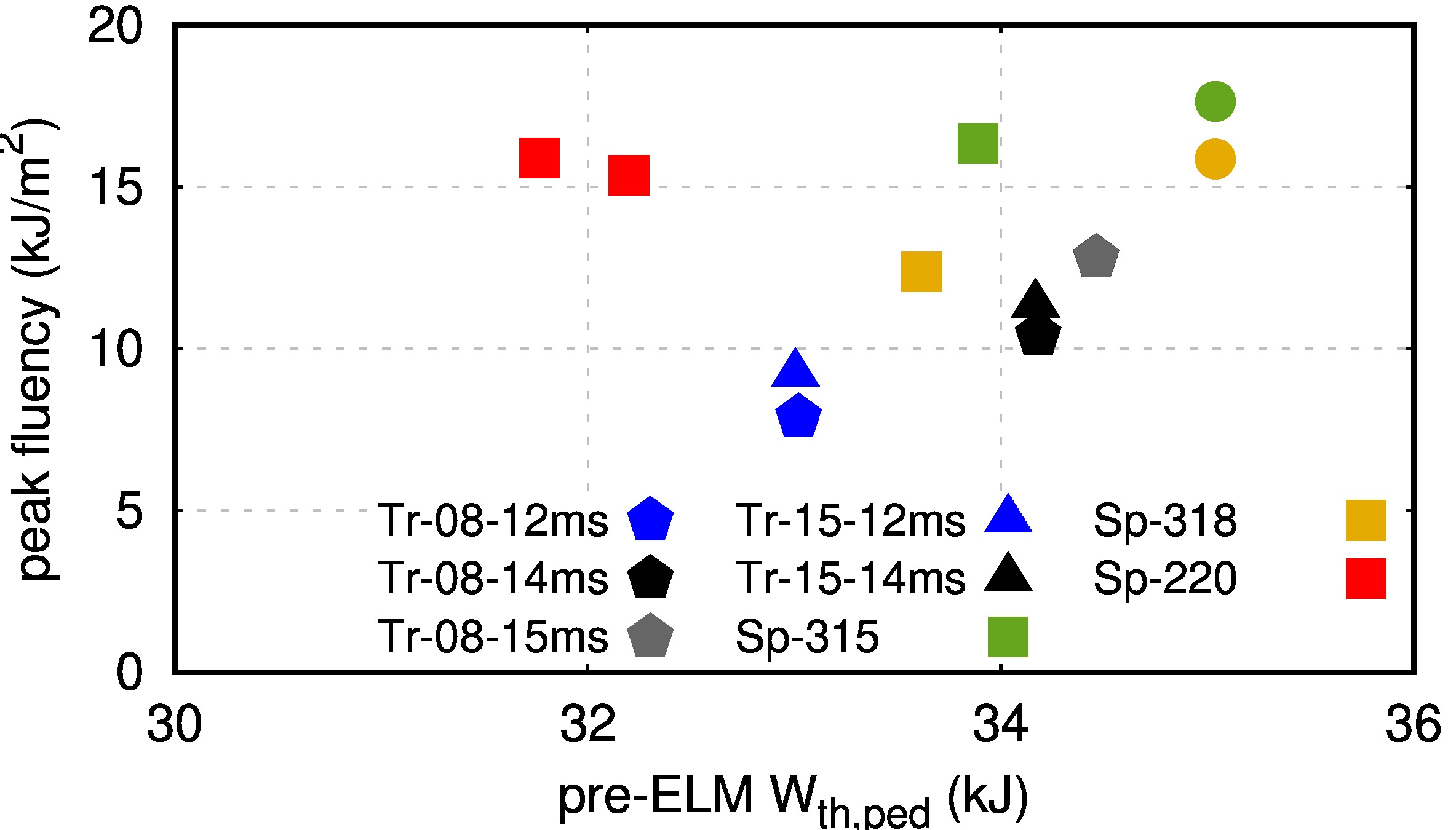}%
    \caption{Peak fluency for spontaneous (in circles and squares) and pellet-triggered ELMs (in pentagons for small pellets and triangles for large pellets). Pellet-triggered ELMs show a moderate decrease in the peak fluency with respect to all but one spontaneous ELMs.}%
    \label{fig:fluencyMax}%
\end{figure}

Comparing the heat flux profiles between different types of ELMs for a single time point clearly evidenced the broader wetted area associated to spontaneous ELMs with respect to pellet-triggered ELMs. However, a more relevant metric is the peak of the time-integrated heat flux deposition profiles. Such time integration is considered through the duration of the ELM crash. This is precisely the target fluency, $\varepsilon$, defined in eqn.~\ref{eq:fluency}. For each of the simulations presented here, but excluding those with an insufficient toroidal resolution (\texttt{Sp-212}), we show the peak target fluency as a function of the pre-ELM pedestal stored thermal energy in fig.~\ref{fig:fluencyMax}. Following the convention from fig.~\ref{fig:compare-energy-losses}, spontaneous ELMs are symbolised with circles (for the first ELM in a series) and with squares (for the subsequent ELMs in a series), and pellet-triggered ELMs are symbolised with pentagons and triangles for pellets with ${0.8\times10^{20}}$ and ${1.5\times10^{20}\mathrm{D}}$ atoms, respectively. 

Four out of five pellet-triggered ELMs show a decrease in the peak target fluency when compared against the converged spontaneous ELMs. In particular, when the ELM crash is triggered significantly early in the ELM cycle the peak fluency can be clearly reduced (\texttt{Tr-08-12ms} and \texttt{Tr-08-14ms}). Furthermore, the pellet size is not observed to play an important role in the peak target fluency. The only pellet-triggered ELM that does not see a reduction of the peak fluency corresponds to pellet injection only one millisecond before the spontaneous ELM would take place, \texttt{Tr-08-15ms}. For this ELM, the pre-ELM pedestal stored thermal energy is comparable to the rest of the spontaneous ELMs. 

\section{Conclusions}\label{conclusions}

We present detailed comparisons between spontaneous and pellet-triggered ELMs simulated with JOREK. The spontaneous ELMs simulated for this study extend recent work of modelling full type-I ELM cycles~\cite{Cathey2020}. Exploiting recent code optimizations, we perform computations at higher toroidal resolution such that simulations are now converged not only in terms of 0D quantities like the energy losses, but also regarding time scales of the crashes and peak divertor heat fluencies. The pellet-triggered ELMs are simulated by introducing pellets at different stages of the pedestal build up. Pellets of two different sizes are considered covering the experimentally accessible range approximately. This novel approach at simulating pellet-triggered ELMs has qualitatively reproduced the lag-time that exists in metal walled devices like AUG-W and JET-ILW as thoroughly described in Ref.~\cite{Futatani2020}. All the simulations presented have been performed with realistic plasma flows (ion velocity comprised of ExB plus diamagnetic flow, but neglecting higher order terms like the polarisation drift). For the spontaneous ELMs, taking into account these realistic plasma flows is a necessary requirement to recover the cyclical dynamics~\cite{Orain2015,Cathey2020}. For the pellet-triggered ELMs, these simulations consider, for the first time, such realistic plasma flows. Additionally, the pellet-triggered ELMs presented here push the previous state-of-the-art further by considering realistic values for the pedestal resistivity.

At first, a detailed analysis of the non-linear dynamics for one spontaneous ELM and one pellet-triggered ELM was presented to describe the similarities and differences between the two events. In particular, the spontaneous ELMs shown in this work feature a precursor phase that lasts roughly one millisecond, while the pellet-triggered ELMs are exited abruptly (${\sim0.1~\mathrm{ms}}$) after the pellet-injection. The pellet-triggered ELM is excited approximately when the pellet has penetrated slightly beyond the maximum pressure gradient region of the pedestal, which is consistent with experimental observations at AUG~\cite{Kocsis_2007}. The pellet-triggered ELMs feature ${\mathrm{n}=1}$ as the dominant toroidal mode number, while the spontaneous ELMs feature peeling-ballooning modes with toroidal mode numbers of ${\mathrm{n}=2,3,4}$ as the dominant modes during the ELM crash. Both results are in agreement with experimental observations~\cite{Poli_2010,Mink2017}. The dominant ${\mathrm{n}=1}$ mode in the pellet-triggered ELM simulations is associated with a helical perturbation directly induced by the pellet injection, and it is shown to have important implications onto the heat flux arriving at the divertor tiles. 

Via a thorough analysis of the heat flux deposition profiles for the different ELMs, it was determined that the wetted area associated with spontaneous ELMs was larger than for pellet-triggered ELMs. An explanation is proposed, by noting that the spontaneous ELMs split the primary strike line onto several secondary strike lines (arriving at different positions in the divertor target), which deposit the thermal energy expelled by the spontaneous ELM onto the area between the primary and secondary strike lines. On the other hand, pellet-triggered ELMs observe only the primary strike line together with a single secondary strike line that is associated with the ${\mathrm{n}=1}$ helical perturbation directly induced by the pellet injection. Most of the energy ejected by the pellet-triggered ELM is then deposited onto the smaller area between the primary and secondary strike lines.

A systematic comparison between several pellet-triggered and spontaneous ELMs is presented showing that large pellets induce energy losses slightly larger than those obtained with smaller pellet injection. We also observe that injecting pellets at earlier times during the pedestal build up leads to smaller energy losses with respect to pellet injections at times closer to the onset of the spontaneous ELMs. In particular, we see that injecting pellets ${\sim4 - 2~\mathrm{ms}}$ before the appearance of the spontaneous ELM leads to a reduction of the ELM energy loss, with respect to the spontaneous ELM, while injecting ${1~\mathrm{ms}}$ before the spontaneous ELM onset leads to comparable energy losses.

Associated to the reduction in the thermal energy losses between spontaneous and pellet-triggered ELMs, the peak divertor incident power is reduced by means of pellet injection. Similarly, and despite the reduction in divertor wetted area related to the pellet-triggered ELMs, the peak fluency that reaches the divertor over the duration of the ELMs is lower for pellet-triggered ELMs with respect to spontaneous ELMs. This difference is most pronounced when comparing the ELMs triggered by pellet injection ${\sim4~\mathrm{ms}}$ before the onset of the spontaneous ELM. 

The work presented here is the first step at simulating pellet-triggered ELMs and comparing triggered and spontaneous ELMs at ASDEX Upgrade. Having shown qualitative agreement on several fronts, future work will be devoted towards producing more quantitative comparisons by injecting pellets onto plasmas with pedestals of different characteristics. Additionally, we aim to investigate pellet pacing by periodic pellet injection into an ELM cycle simulation with a higher frequency than the natural ELM frequency.

\section*{Acknowledgements}

The authors would like to thank Thomas Eich and Davide Silvagni for fruitful discussions. This work has been carried out within the framework of the EUROfusion Consortium and has received funding from the Euratom research and training program 2014-2018 and 2019-2020 under grant agreement No 633053. The views and opinions expressed herein do not necessarily reflect those of the European Commission. In particular, contributions by EUROfusion work packages Enabling Research (EnR) and Medium Size Tokamaks (MST) is acknowledged. We acknowledge PRACE for awarding us access to MareNostrum at Barcelona Supercomputing Center (BSC), Spain. Some of the simulations were performed using the Marconi-Fusion supercomputer. 

\bibliography{main}

\begin{thebibliography}{10}

\bibitem{Loarte2014}
A~Loarte, G~Huijsmans, S.~Futatani, L.R. Baylor, T.E. Evans, D.~M. Orlov,
  O.~Schmitz, M.~Becoulet, P.~Cahyna, Y.~Gribov, A.~Kavin, A.~Sashala Naik,
  D.J. Campbell, T.~Casper, E.~Daly, H.~Frerichs, A.~Kischner, R.~Laengner,
  S.~Lisgo, R.A. Pitts, G.~Saibene, and A.~Wingen.
\newblock Progress on the application of {ELM} control schemes to {ITER}
  scenarios from the non-active phase to {DT} operation.
\newblock {\em Nuclear Fusion}, 54(3):033007, 2014.

\bibitem{Labit_2019}
B.~Labit, T.~Eich, G.F. Harrer, E.~Wolfrum, M.~Bernert, M.G. Dunne,
  L.~Frassinetti, P.~Hennequin, R.~Maurizio, A.~Merle, et~al.
\newblock Dependence on plasma shape and plasma fueling for small
  edge-localized mode regimes in {TCV} and {ASDEX} upgrade.
\newblock {\em Nuclear Fusion}, 59(8):086020, jun 2019.

\bibitem{Viezzer_2018}
E.~Viezzer.
\newblock Access and sustainment of naturally {ELM}-free and small-{ELM}
  regimes.
\newblock {\em Nuclear Fusion}, 58(11):115002, sep 2018.

\bibitem{Evans2008}
T.E. Evans, M.E. Fenstermacher, R.A. Moyer, T.H. Osborne, J.G. Watkins,
  P.~Gohil, I.~Joseph, M.J. Schaffer, L.R. Baylor, M.~Bécoulet, J.A. Boedo,
  K.H. Burrell, J.S. deGrassie, K.H. Finken, T.~Jernigan, M.W. Jakubowski, C.J.
  Lasnier, M.~Lehnen, A.W. Leonard, J.~Lonnroth, E.~Nardon, V.~Parail,
  O.~Schmitz, B.~Unterberg, and W.P. West.
\newblock Rmp {ELM} suppression in diii-d plasmas with iter similar shapes and
  collisionalities.
\newblock {\em Nuclear Fusion}, 48(2):024002, 2008.

\bibitem{Lang2004_1}
P.T. Lang, G.D. Conway, T.~Eich, L.~Fattorini, O.~Gruber, S.~Günter, L.D.
  Horton, S.~Kalvin, A.~Kallenbach, M.~Kaufmann, G.~Kocsis, A.~Lorenz, M.E.
  Manso, M.~Maraschek, V.~Mertens, J.~Neuhauser, I.~Nunes, W.~Schneider,
  W.~Suttrop, H.~Urano, and {the ASDEX Upgrade Team}.
\newblock {ELM} pace making and mitigation by pellet injection in {ASDEX
  Upgrade}.
\newblock {\em Nuclear Fusion}, 44(5):685, 2004.

\bibitem{Baylor2013PRL}
L.~R. Baylor, N.~Commaux, T.~C. Jernigan, N.~H. Brooks, S.~K. Combs, T.~E.
  Evans, M.~E. Fenstermacher, R.~C. Isler, C.~J. Lasnier, S.~J. Meitner, R.~A.
  Moyer, T.~H. Osborne, P.~B. Parks, P.~B. Snyder, E.~J. Strait, E.~A.
  Unterberg, , and A.~Loarte.
\newblock Reduction of edge-localized mode intensity using high-repetition-rate
  pellet injection in tokamak {H-Mode} plasmas.
\newblock {\em Physical Review Letters}, 110:245001, 2013.

\bibitem{Wenninger2011}
R~P Wenninger, T~H Eich, G~T~A Huysmans, P~T Lang, S~Devaux, S~Jachmich,
  F~Köchl, and {JET EFDA Contributors}.
\newblock Scrape-off layer heat transport and divertor power deposition of
  pellet-induced edge localized modes.
\newblock {\em Plasma Physics and Controlled Fusion}, 53(10):105002, aug 2011.

\bibitem{Huysmans2007}
G.T.A. Huysmans and O.~Czarny.
\newblock {MHD} stability in x-point geometry: simulation of {ELMs}.
\newblock {\em Nuclear Fusion}, 47(7):659, 2007.

\bibitem{Czarny2008}
Olivier Czarny and Guido Huysmans.
\newblock Bezier surfaces and finite elements for {MHD} simulations.
\newblock {\em Journal of Computational Physics}, 227(16):7423 -- 7445, 2008.

\bibitem{Hoelzl2021}
M~Hoelzl, GTA Huijsmans, SJP Pamela, M~Becoulet, E~Nardon, FJ~Artola, B~Nkonga,
  CV~Atanasiu, V~Bandaru, A~Bhole, D~Bonfiglio, A~Cathey, O~Czarny, A~Dvornova,
  T~Feher, A~Fil, E~Franck, S~Futatani, M~Gruca, H~Guillard, JW~Haverkort,
  I~Holod, D~Hu, SK~Kim, SQ~Korving, L~Kos, I~Krebs, L~Kripner, G~Latu, F~Liu,
  P~Merkel, D~Meshcheriakov, V~Mitterauer, S~Mochalskyy, JA~Morales, R~Nies,
  N~Nikulsin, F~Orain, D~Penko, J~Pratt, R~Ramasamy, P~Ramet, C~Reux,
  N~Schwarz, P~Singh Verma, SF~Smith, C~Sommariva, E~Strumberger, DC~vanVugt,
  M~Verbeek, E~Westerhof, F~Wieschollek, and J~Zielinski.
\newblock The jorek non-linear extended mhd code and applications to
  large-scale instabilities and their control in magnetically confined fusion
  plasmas.
\newblock 2020.

\bibitem{Cathey2020}
A.~Cathey, M.~Hoelzl, K.~Lackner, G.T.A. Huijsmans, M.G. Dunne, E.~Wolfrum,
  S.J.P. Pamela, F.~Orain, and S.~Günter.
\newblock Non-linear extended {MHD} simulations of type-{I} edge localised mode
  cycles in {ASDEX Upgrade} and their underlying triggering mechanism.
\newblock {\em Nuclear Fusion}, 60(12):124007, nov 2020.

\bibitem{Lang_2014}
P.T. Lang, A.~Burckhart, M.~Bernert, L.~Casali, R.~Fischer, O.~Kardaun,
  G.~Kocsis, M.~Maraschek, A.~Mlynek, B.~Plöckl, M.~Reich, F.~Ryter,
  J.~Schweinzer, B.~Sieglin, W.~Suttrop, T.~Szepesi, G.~Tardini, E.~Wolfrum,
  D.~Zasche, and H.~Zohm and.
\newblock {ELM} pacing and high-density operation using pellet injection in the
  {ASDEX Upgrade} all-metal-wall tokamak.
\newblock {\em Nuclear Fusion}, 54(8):083009, jun 2014.

\bibitem{Futatani2020}
S.~Futatani, A.~Cathey, M.~Hoelzl, and {et al}.
\newblock Transition from no-{ELM} response to pellet {ELM} triggering during
  pedestal build-up -- insights from extended {MHD} simulations.
\newblock {\em Nuclear Fusion}, page submitted, 2020.

\bibitem{Kocsis_2007}
G.~Kocsis, S.~K{\'{a}}lvin, P.T. Lang, M.~Maraschek, J.~Neuhauser,
  W.~Schneider, and T.~Szepesi.
\newblock Spatio-temporal investigations on the triggering of pellet induced
  {ELMs}.
\newblock {\em Nuclear Fusion}, 47(9):1166--1175, aug 2007.

\bibitem{Romanelli2009}
F.~Romanelli, R.~Kamendje, and {on behalf of JET-EFDA Contributors}.
\newblock Overview of {JET} results.
\newblock {\em Nuclear Fusion}, 49(10):104006, 2009.

\bibitem{Lang2008}
P.T. Lang, K.~Lackner, M.~Maraschek, B.~Alper, E.~Belonohy, K.~Gál, J.~Hobirk,
  A.~Kallenbach, S.~Kálvin, G.~Kocsis, C.P.~Perez von Thun, W.~Suttrop,
  T.~Szepesi, R.~Wenninger, H.~Zohm, {the ASDEX Upgrade Team}, and {JET-EFDA
  contributors}.
\newblock Investigation of pellet-triggered {MHD} events in {ASDEX Upgrade} and
  {JET}.
\newblock {\em Nuclear Fusion}, 48(9):095007, 2008.

\bibitem{Lang_2013}
P.T. Lang, D.~Frigione, A.~G{\'{e}}raud, T.~Alarcon, P.~Bennett, G.~Cseh,
  D.~Garnier, L.~Garzotti, F.~Köchl, G.~Kocsis, M.~Lennholm, R.~Neu,
  R.~Mooney, S.~Saarelma, and B.~Sieglin and.
\newblock {ELM} pacing and trigger investigations at {JET} with the new
  {ITER}-like wall.
\newblock {\em Nuclear Fusion}, 53(7):073010, may 2013.

\bibitem{Lang2015}
P~T Lang, H~Meyer, G~Birkenmeier, A~Burckhart, I~S Carvalho, E~Delabie,
  L~Frassinetti, G~Huijsmans, A~Loarte G~Kocsi~and, C~F Maggi, M~Maraschek,
  B~Ploeckl, F~Rimini, F~Ryter, S~Saarelma, T~Szepesi, E~Wolfrum, {ASDEX
  Upgrade Team}, and {JET Contributors}.
\newblock {ELM} control at the {L-H} transition by means of pellet pacing in
  the asdex upgrade and {JET} all-metal-wall tokamaks.
\newblock {\em Plasma Physics and Controlled Fusion}, 57(4):045011, 2015.

\bibitem{Hoelzl2020A}
M.~Hoelzl, D.~Hu, E.~Nardon, and G.~T.~A. Huijsmans.
\newblock First predictive simulations for deuterium shattered pellet injection
  in {ASDEX Upgrade}.
\newblock {\em Physics of Plasmas}, 27(2):022510, 2020.

\bibitem{Orain2013}
F.~Orain, M.~B\'ecoulet, G.~Dif-Pradalier, G.~Huijsmans, S.~Pamela, E.~Nardon,
  C.~Passeron, G.~Latu, V.~Grandgirard, A.~Fil, A.~Ratnani, I.~Chapman,
  A.~Kirk, A.~Thornton, M.~Hoelzl, and P.~Cahyna.
\newblock Non-linear magnetohydrodynamic modeling of plasma response to
  resonant magnetic perturbations.
\newblock {\em Physics of Plasmas}, 20(10):102510, 2013.

\bibitem{Pamela2017}
S.J.P. Pamela, G.T.A. Huijsmans, T.~Eich, S.~Saarelma, I.~Lupelli, C.F. Maggi,
  C.~Giroud, I.T. Chapman, S.F. Smith, L.~Frassinetti, M.~Becoulet, M.~Hoelzl,
  F.~Orain, S.~Futatani, and JET Contributors.
\newblock Recent progress in the quantitative validation of {JOREK} simulations
  of {ELMs} in {JET}.
\newblock {\em Nuclear Fusion}, 57(7):076006, 2017.

\bibitem{Groebner_Burrell_Seraydarian_1990}
R.~J. Groebner, K.~H. Burrell, and R.~P. Seraydarian.
\newblock Role of edge electric field and poloidal rotation in the {L-H}
  transition.
\newblock {\em Physical Review Letters}, 64(25):3015–3018, Jun 1990.

\bibitem{cavedon2017pedestal}
M~Cavedon, T~P{\"u}tterich, Eleonora Viezzer, FM~Laggner, A~Burckhart, M~Dunne,
  R~Fischer, A~Lebschy, F~Mink, U~Stroth, et~al.
\newblock Pedestal and {Er} profile evolution during an edge localized mode
  cycle at {ASDEX Upgrade}.
\newblock {\em Plasma Physics and Controlled Fusion}, 59(10):105007, 2017.

\bibitem{rogers1999diamagnetic}
BN~Rogers and JF~Drake.
\newblock Diamagnetic stabilization of ideal ballooning modes in the edge
  pedestal.
\newblock {\em Physics of Plasmas}, 6(7):2797--2801, 1999.

\bibitem{Sauter1999}
O.~Sauter, C.~Angioni, and Y.~R. Lin-Liu.
\newblock Neoclassical conductivity and bootstrap current formulas for general
  axisymmetric equilibria and arbitrary collisionality regime.
\newblock {\em Physics of Plasmas}, 6(7):2834--2839, 1999.

\bibitem{Sauter2002}
O.~Sauter, C.~Angioni, and Y.~R. Lin-Liu.
\newblock Neoclassical conductivity and bootstrap current formulas for general
  axisymmetric equilibria and arbitrary collisionality regime.
\newblock {\em Physics of Plasmas}, 6(7):2834--2839, 1999.

\bibitem{Futatani2014}
S.~Futatani, G.~Huijsmans, A.~Loarte, L.R. Baylor, N.~Commaux, T.C. Jernigan,
  M.E. Fenstermacher, C.~Lasnier, T.H. Osborne, and B.~Pegourié.
\newblock Non-linear {MHD} modelling of {ELM} triggering by pellet injection in
  {DIII-D} and implications for {ITER}.
\newblock {\em Nuclear Fusion}, 54(7):073008, 2014.

\bibitem{Nardon2007}
E.~Nardon, M.~B\'ecoulet, G.~Huysmans, and O.~Czarny.
\newblock Magnetohydrodynamics modelling of {H}-mode plasma response to
  external resonant magnetic perturbations.
\newblock {\em Physics of Plasmas}, 14(9):092501, 2007.

\bibitem{Orain2017}
F.~Orain, M.~Hoelzl, E.~Viezzer, M.~Dunne, M.~Becoulet, P.~Cahyna, G.T.A.
  Huijsmans, J.~Morales, M.~Willensdorfer, W.~Suttrop, A.~Kirk, S.~Pamela,
  S.~Guenter, K.~Lackner, E.~Strumberger, A.~Lessig, {the ASDEX Upgrade Team},
  and {the EUROfusion MST1 Team}.
\newblock Non-linear modeling of the plasma response to {RMPs} in {ASDEX
  Upgrade}.
\newblock {\em Nuclear Fusion}, 57(2):022013, 2017.

\bibitem{Becoulet2014}
M.~B\'ecoulet, F.~Orain, G.~T.~A. Huijsmans, S.~Pamela, P.~Cahyna, M.~Hoelzl,
  X.~Garbet, E.~Franck, E.~Sonnendr\"ucker, G.~Dif-Pradalier, C.~Passeron,
  G.~Latu, J.~Morales, E.~Nardon, A.~Fil, B.~Nkonga, A.~Ratnani, and
  V.~Grandgirard.
\newblock Mechanism of edge localized mode mitigation by resonant magnetic
  perturbations.
\newblock {\em Phys. Rev. Lett.}, 113:115001, Sep 2014.

\bibitem{LiuF2015}
F.~Liu, G.T.A. Huijsmans, A.~Loarte, A.M. Garofalo, W.M. Solomon, P.B. Snyder,
  M.~Hoelzl, and L.~Zeng.
\newblock Nonlinear {MHD} simulations of {Quiescent H-mode} plasmas in
  {DIII-D}.
\newblock {\em Nuclear Fusion}, 55(11):113002, 2015.

\bibitem{Artola2018}
F.J. Artola, G.T.A. Huijsmans, M.~Hoelzl, P.~Beyer, A.~Loarte, and Y.~Gribov.
\newblock Non-linear magnetohydrodynamic simulations of edge localised mode
  triggering via vertical position oscillations in {ITER}.
\newblock {\em Nuclear Fusion}, 58(9):096018, jul 2018.

\bibitem{Huysmans2009}
G~T~A Huysmans, S~Pamela, E~van~der Plas, and P~Ramet.
\newblock Non-linear {MHD} simulations of edge localized modes ({ELMs}).
\newblock {\em Plasma Physics and Controlled Fusion}, 51(12):124012, 2009.

\bibitem{Futatani2019}
S.~Futatani, S.~Pamela, L.~Garzotti, G.T.A. Huijsmans, M.~Hoelzl, D.~Frigione,
  M.~Lennholm, and and.
\newblock Non-linear magnetohydrodynamic simulations of pellet triggered
  edge-localized modes in {JET}.
\newblock {\em Nuclear Fusion}, 60(2):026003, dec 2019.

\bibitem{Huijsmans2015}
G.~T.~A. Huijsmans, C.~S. Chang, N.~Ferraro, L.~Sugiyama, F.~Waelbroeck, X.~Q.
  Xu, A.~Loarte, and S.~Futatani.
\newblock Modelling of edge localised modes and edge localised mode control.
\newblock {\em Physics of Plasmas}, 22(2):021805, 2015.

\bibitem{Hoelzl2012A}
M.~Hoelzl, S.~Guenter, R.~P. Wenninger, W.-C. Mueller, G.~T.~A. Huysmans,
  K.~Lackner, and I.~Krebs.
\newblock Reduced-magnetohydrodynamic simulations of toroidally and poloidally
  localized edge localized modes.
\newblock {\em Physics of Plasmas}, 19(8):082505, 2012.

\bibitem{Krebs2013}
I.~Krebs, M.~Hoelzl, K.~Lackner, and S.~Guenter.
\newblock Nonlinear excitation of low-n harmonics in reduced
  magnetohydrodynamic simulations of edge-localized modes.
\newblock {\em Physics of Plasmas}, 20(8):082506, 2013.

\bibitem{Hoelzl2018}
M~Hoelzl, G~T~A Huijsmans, F~Orain, F~J Artola, S~Pamela, M~Becoulet, D~{van
  Vugt}, F~Liu, S~Futatani, A~Lessig, E~Wolfrum, F~Mink, E~Trier, M~Dunne,
  E~Viezzer, T~Eich, B~Vanovac, L~Frassinetti, S~Guenter, K~Lackner, I~Krebs,
  {ASDEX Upgrade Team}, and {EUROfusion MST1 Team}.
\newblock Insights into {type-I} {ELMs} and {ELM} control methods from {JOREK}
  {MHD} simulations.
\newblock {\em Contributions to Plasma Physics}, 58:518, 2018.

\bibitem{Orain2019}
F.~Orain, M.~Hoelzl, F.~Mink, M.~Willensdorfer, M.~Bécoulet, M.~Dunne,
  S.~Günter, G.~Huijsmans, K.~Lackner, S.~Pamela, W.~Suttrop, and E.~Viezzer.
\newblock Non-linear modeling of the threshold between {ELM} mitigation and
  {ELM} suppression by resonant magnetic perturbations in {ASDEX Upgrade}.
\newblock {\em Physics of Plasmas}, 26(4):042503, 2019.

\bibitem{Lang2003}
P.~T. Lang, P.~Cierpka, O.~Gehre, M.~Reich, C.~Wittmann, A.~Lorenz,
  D.~Frigione, S.~Kalvin, G.~Kocsis, and S.~Maruyama.
\newblock A system for cryogenic hydrogen pellet high speed inboard launch into
  a fusion device via guiding tube transfer.
\newblock {\em Review of Scientific Instruments}, 74(9):3974--3983, 2003.

\bibitem{Holod2020}
I~Holod, M~Hoelzl, PS~Verma, GTA Huijsmans, R~Nies, and {JOREK Team}.
\newblock New developments regarding the jorek solver and physics based
  preconditioner.
\newblock {\em Journal of Computational Physics}, submitted.

\bibitem{Eich2005}
T~Eich, A~Herrmann, J~Neuhauser, R~Dux, J~C Fuchs, S~Günter, L~D Horton,
  A~Kallenbach, P~T Lang, C~F Maggi, M~Maraschek, V~Rohde, W~Schneider, and
  {the ASDEX Upgrade Team}.
\newblock Type-i {ELM} substructure on the divertor target plates in {ASDEX
  Upgrade}.
\newblock {\em Plasma Physics and Controlled Fusion}, 47(6):815, 2005.

\bibitem{Orain2015}
F.~Orain, M.~B\'ecoulet, G.~T.~A. Huijsmans, G.~Dif-Pradalier, M.~Hoelzl,
  J.~Morales, X.~Garbet, E.~Nardon, S.~Pamela, C.~Passeron, G.~Latu, A.~Fil,
  and P.~Cahyna.
\newblock Resistive reduced {MHD} modeling of multi-edge-localized-mode cycles
  in tokamak $x$-point plasmas.
\newblock {\em Phys. Rev. Lett.}, 114:035001, Jan 2015.

\bibitem{Jakubowski2009}
M.W. Jakubowski, T.E. Evans, M.E. Fenstermacher, M.~Groth, C.J. Lasnier, A.W.
  Leonard, O.~Schmitz, J.G. Watkins, T.~Eich, W.~Fundamenski, R.A. Moyer, R.C.
  Wolf, L.B. Baylor, J.A. Boedo, K.H. Burrell, H.~Frerichs, J.S. deGrassie,
  P.~Gohil, I.~Joseph, S.~Mordijck, M.~Lehnen, C.C. Petty, R.I. Pinsker,
  D.~Reiter, T.L. Rhodes, U.~Samm, M.J. Schaffer, P.B. Snyder, H.~Stoschus,
  T.~Osborne, B.~Unterberg, E.~Unterberg, and W.P. West.
\newblock Overview of the results on divertor heat loads in {RMP} controlled
  h-mode plasmas on {DIII}-d.
\newblock {\em Nuclear Fusion}, 49(9):095013, aug 2009.

\bibitem{Eich2017A}
T.~Eich, B.~Sieglin, A.J. Thornton, M.~Faitsch, A.~Kirk, A.~Herrmann, and
  W.~Suttrop.
\newblock {ELM} divertor peak energy fluence scaling to iter with data from
  jet, mast and {ASDEX Upgrade}.
\newblock {\em Nuclear Materials and Energy}, 12:84, 2017.

\bibitem{Poli_2010}
F.M. Poli, P.T. Lang, S.E. Sharapov, B.~Alper, and H.R.~Koslowski and.
\newblock Spectra of magnetic perturbations triggered by pellets in {JET}
  plasmas.
\newblock {\em Nuclear Fusion}, 50(2):025004, jan 2010.

\bibitem{Mink2017}
F.Mink, M.Hoelzl, E.Wolfrum, F.Orain, M.Dunne, A.Lessig, S.Pamela, P.Manz,
  M.Maraschek, G.T.A.Huijsmans, M.Becoulet, F.M.Laggner, M.Cavedon, K.Lackner,
  S.Guenter, U.Stroth, and {the ASDEX Upgrade Team}.
\newblock Nonlinear coupling induced toroidal structure of edge localized
  modes.
\newblock {\em Nuclear Fusion}, 58:026011, 2018.

\end{thebibliography}

\end{document}